# Dynamic Range Beyond Bit Depth of a CMOS Image Sensor Using Interleaved Row Readout

Supriyo Ghosh,[1] ⋆ William Martin,[1] Hugh R. A. Jones, [1] Angaraj Duara, [2] Konstantin Stefanov, [2] and Jesper Skottfelt [2]

[1] *Centre for Astrophysics Research, Department of Physics, Astronomy and Mathematics, University of Hertfordshire, Hatfield, Hertfordshire AL10 9AB, UK*

[2] *Centre for Electronic Imaging, School of Physical Sciences, The Open University, Walton Hall, Milton Keynes, MK7 6AA, UK*

**ABSTRACT**

The dynamic range available from a sensor is vital to its utility. The limits on the dynamic range that can be obtained from an image sensor are set by the brightest and faintest objects that can be detected. In recent years, CMOS (complementary metal-oxide-semiconductor) image sensors (CIS) have gained high popularity due to their low cost and high availability. However, as with all detectors, the dynamic range is constrained by the sensor's bit depth. Here, we have modified the readout scheme of a commercial CIS120 sensor from Teledyne e2v, to enhance its dynamic range. We have advanced the interleaved row readout method proposed by Wocial et al. by using a more sophisticated approach, which enables us to readout chosen rows much more frequently to avoid saturation and then readout other rows on the sensor once to form the image. Our laboratory tests provide a dynamic range of 134 dB elevated from the sensor's native 12-bit of about 71 dB. We also built a camera housing that enabled first-time operation of interleaved row readout on-sky to observe the bright stars, Vega and Polaris. In complex mode we obtained unsaturated single exposure images of these bright stars, which have magnitudes near zero and detect background stars with Gaia G magnitudes around 15 in a single exposure, with a detection threshold of $5\sigma$. The achievable dynamic range with this interleaved row readout is limited only by the readout noise and scattering in the camera optics.

**Key words:** Instrument – Astronomical camera – CMOS – High dynamic range – Row readout sequencing – Imaging

Wocial et al by using more sophisticated readout

## 1 INTRODUCTION

In observational astronomy, detectors play a crucial role to record radiation coming from celestial objects. The advancement of detector technology helps us to explore our cosmos in greater detail. In general, astronomical detectors are expected to have several key properties, including but not limited to fast response time, a wide dynamic range, high quantum efficiency, readout speed and fill factor, and small pixel size with uniform linear response. In addition, space applications and astronomy imaging (Kenworthy & Haffert 2025) demand the requirement for low-noise (low background noise such as read noise and dark current) performance. Noise not only sets up the signal-to-noise ratio (SNR) and the dynamic range (DR), but also plays an important role to define the minimum input signal needed for acceptable output signal (Fossum et al. 2024).

Charge-coupled devices (CCDs, Boyle & Smith 1970) have long been used as the detector of choice in optical astronomy because of their excellent performance in terms of linearity, quantum efficiency and fast response time over previous generations of detectors such as photographic plates. In recent years, the usage of CCDs has been reduced to some extent in applications of time domain astronomy and/or objects in motion due to their limited readout speed and rise of alternative technologies, for example, complementary metal-oxide-semiconductor (CMOS) image sensors (CIS) (Fossum 1997). Moreover, CIS offer additional advantages such as individual pixel signal processing, and lower voltage operation, power consumption and cost over traditional CCD-based imagers (Bigas et al. 2006).

The performance of any sensor relies on the array of photo-detectors and associated readout circuits. In general, the pixel defines the sensitivity, spectral response, and full well capacity (FWC) of the image sensor, while its associated circuitry determines the noise, resolution, readout speed, and sampling scheme. Achieving high linearity and SNR at low power consumption, in addition to a high DR, are general requirements for readout circuits used in these image sensors (Kavusi et al. 2006b). DR of an image sensor can be defined as the flux ratio between the brightest and faintest signals that are possible to measure accurately. DR is primarily determined by FWC of a pixel and the noise characteristics of the readout.

In the early days of CIS, their applications to higher resolution astronomical observations was restricted due to the lower DR, poorer linearity, pixel non-uniformity and fill factor in comparison to CCD (Hoffman et al. 2005; Bigas et al. 2006). In addition, the conventional CIS was front-illuminated and thus, they have worse quantum efficiency (QE) compared to CCDs. With the rapid advancement of CMOS technology, the new generation of sensors, such as scientific CMOS (sCMOS) was introduced in 2009 by Andor Technology, Fairchild Imaging (BAE Systems) in collaboration with PCO Imaging. Such sensors offer low noise level, faster readout, reasonable pixel and sensor sizes, high-cadence subframe readout and built-in anti-blooming (Qiu et al. 2013; Pratlong et al. 2016; Earnshaw et al. 2022; Alarcon et al. 2023; and Greffe et al. 2022 for a review) and are now being routinely used (e.g., Apergis et al. 2025). Furthermore, using skipper technology, the read noise can be reduced to deep sub-electron noise level (~ $0.15e^-$, Stefanov et al. 2020; Lapi et al. 2024). In addition, a recent advancement to back-illuminated CIS has been demonstrated a capability of reaching comparable QE (~ 95% at optical wavelengths, Setälä et al. 2023) to CCDs. Hence, back-illuminated sCMOS begins to be suitable and increasingly popular in application of astronomy for their greater performance features despite growing camera design complexity.

⋆ E-mail: s.ghosh3@herts.ac.uk

Nevertheless, in astronomy, high dynamic range (HDR) is essential for imaging both extremely bright and faint objects simultaneously within a single frame. This capability enables the observation of features such as the galactic centre alongside its faint outer arms, direct imaging of exoplanets, and accurate calibration using both strong and weak spectral lines in spectroscopy. Without HDR techniques, imaging at the standard dynamic range corresponding to the bit depth can cause bright objects to become saturated (overexposed), while faint objects may be lost in noise under low-light conditions. This leads to distorted features and a degradation of overall image quality.The problems of HDR-based object detection and exposure adaption technologies are manifest in a number of areas beyond astronomy. This a particular problem where illumination conditions change rapidly. This includes a wide range of fields such as microscopy, machine vision, driving in direct sunlight, surveillance, and autonomous driving (e.g., Paul & Chung 2018; Mukherjee et al. 2021; Onzon et al. 2021; Eom & Kim 2025).

Over the years, various techniques have been developed to enhance DR (see Chen et al. 2025 for a review). These approaches include, but are not limited to, modifications to the structure of pixels to extend FWC and/or reduce noise levels, as well as multiple exposure scheme. Examples of such methods include changing the well capacity by adjusting the resetting signal one or more times during integration (Decker et al. 1998); dual-gain by implementing two amplifiers in each pixel to operate at different gain settings and read-noise levels (Earnshaw et al. 2022); multiple-gain by modulating charge storage capacity using additional switches and capacitors in each pixel (Huggett et al. 2009; Ma et al. 2017; Takayanagi et al. 2018); lateral overflow integration capacitor (LOFIC) by adding an extra capacitor to capture excess charge and prevent sensor saturation (Murata et al. 2020); logarithmic pixel response by incorporating a logarithmic compressing transistor in each pixel (Burghartz et al. 2006); and linear-logarithmic charge compensation, where pixels operate in either linear or logarithmic mode depending on incident light intensity (Li et al. 2016). All these schemes aim to enhance pixel FWC. The LOFIC scheme has achieved a DR of 130 dB (Murata et al. 2020), while a DR of over 90 dB has been reported in dual-gain mode by Takayanagi et al. (2018). A DR of about 161 dB was achieved by the linear-logarithmic charge compensation method (Li et al. 2016).

Multiple capturing schemes include obtaining multiple frames at different exposure times to accommodate both bright and faint objects (Yadid-Pecht & Fossum 1997; Mase et al. 2005; Sasaki et al. 2007; Yamada et al. 2008), reading out non-destructively during integration without affecting the accumulated signal (Yang et al. 1999; Bidermann et al. 2003), and time-to-saturation readout, which involves implementing a voltage comparator and an analog memory in each active pixel to convert each photocurrent signal into the integration time needed to reach saturation (Stoppa et al. 2002). While the multiple exposure scheme for bright and faint objects has achieved a DR of 140 dB (Yamada et al. 2008), DRs of about 120 dB and 132 dB have been achieved by non-destructive readout (Kavusi et al. 2006b) and time-to-saturation readout schemes (Stoppa et al. 2002), respectively.

Other approaches include self-reset pixels, which can operate either synchronously (Bermak et al. 2002) or asynchronously (Lenero-Bardallo et al. 2017), based on predefined reference voltage(s). A DR of 161 dB was achieved with the synchronous self-reset approach (Kavusi et al. 2006b), while the asynchronous scheme provides a DR of up to 130 dB. In addition, Kavusi et al. (2006a) implemented the folded multiple capture scheme, which combines synchronous self-reset with multiple capture, achieving a DR of 138 dB. A modern advanced technique of 3D stacking architecture was introduced by Hirata et al. (2021), where the pixel array was segmented into multiple blocks and exposure time in each block can be controlled independently. A DR of 134 dB can be achieved by this architecture.

Wocial et al. (2022) implemented an interlaced row-wise readout multiple-exposure scheme. In this scheme, rows could be selected based on the brightness level of the object for reading out single or multiple times. The user-defined bright rows read out multiple times, while faint rows read out once. An increase of 34 dB DR was presented using this scheme with an increment of 17 dB peak SNR. This scheme is relatively simple without the requirement of any additional circuitry, however, has not yet been implemented for on-sky observations. Additionally, more advanced and specialized CIS versions are now available on the market, offering larger buffer sizes, higher bit depths, and enhanced processing capabilities. With these advancements, integrating this readout option into the controller interface is becoming increasingly important. Testing and improving such controller interfaces is a key part of our development project, along with implementing such readout technique both in the laboratory and in real on-sky observations.

In this work, we used a back-illuminated commercial CMOS sensor, CIS120 from Teledyne e2v. CIS120 is specifically designed for space applications with arbitrary row readout and is capable of operating at wavelengths from X-rays to infrared. It features an array of 2048 × 2048 pixels with a 10 $\mu$m pixel pitch and offers a typical quantum efficiency of 90% at 600 nm. The sensor can operate in either rolling shutter mode (reading out pixels line by line sequentially) or global shutter mode (capturing the signal in all pixels simultaneously), with flexible resolution settings from 8 to 14 bits and different analog-to-digital converter (ADC) gain settings, specifically high and low, to optimize performance in different lighting conditions. A bit depth of 12 was utilized in this study. We refer to Te2v[1] for additional details.

In this study, we enhanced the DR of the commercial CIS120 sensor by implementing a more sophisticated version of the interleaved row readout method (Wocial et al. 2022). We refer to this updated version of the interleaved row readout, as applied to CIS120, as complex mode to distinguish it from the standard readout mode, which serves as our baseline for comparison in this paper. This complex mode incorporates a larger number of mini-frames and utilizes the split. The structure of the paper is as follows: Section 2 describes the experimental setup. Section 3 provides a comprehensive characterization of the CIS120 sensor. Section 4 explains how the DR can be enhanced for the CIS120 and presents results from both laboratory and on-sky experiments. Section 5 discusses the various trade-offs made during our experiment, and Section 6 summarizes our study.

[1] https://www.teledynespaceimaging.com/en-us/products/cmos-general-purpose-sensors/

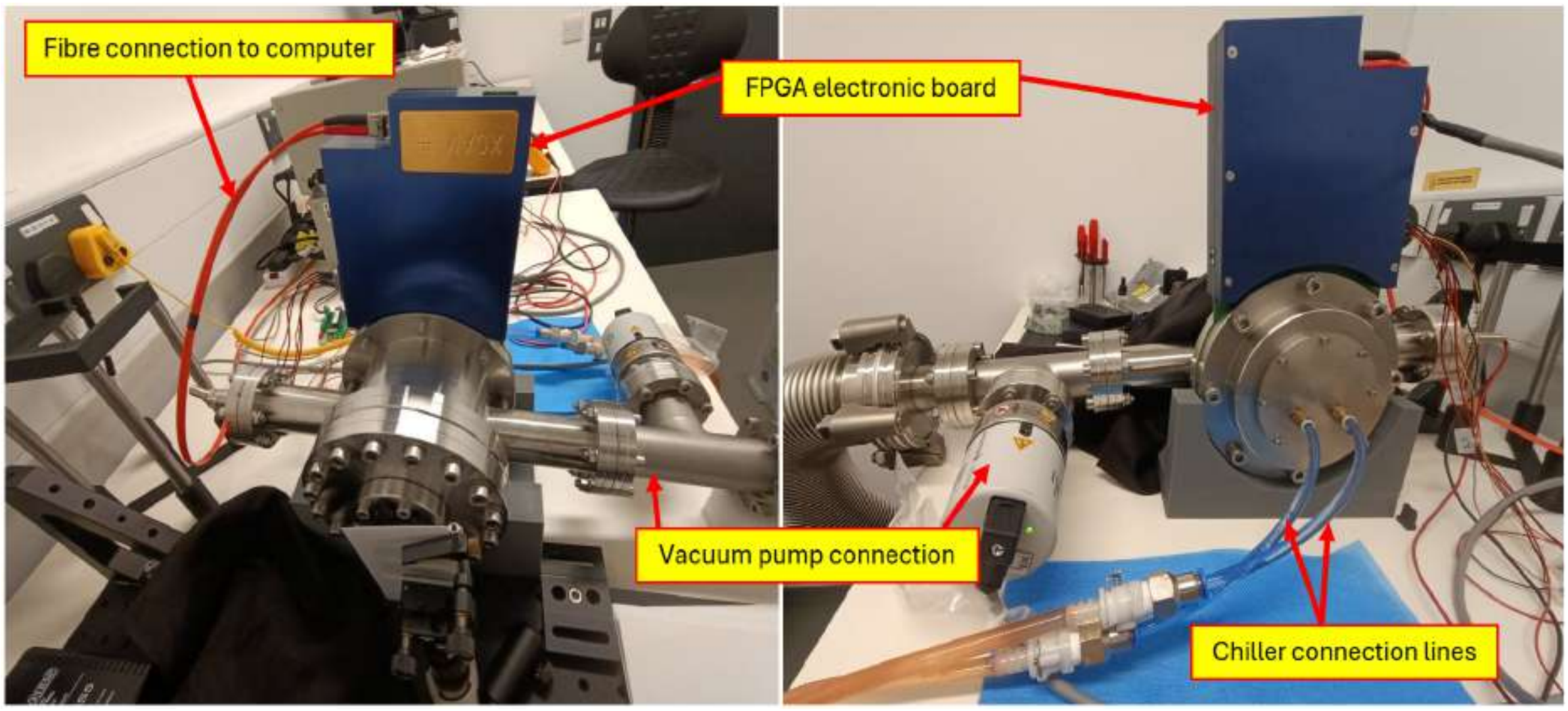

**Figure 1.** The experimental setup used for the CIS120 investigation in the laboratory is shown. The left image shows the front side of the camera, while the right image displays the back side.

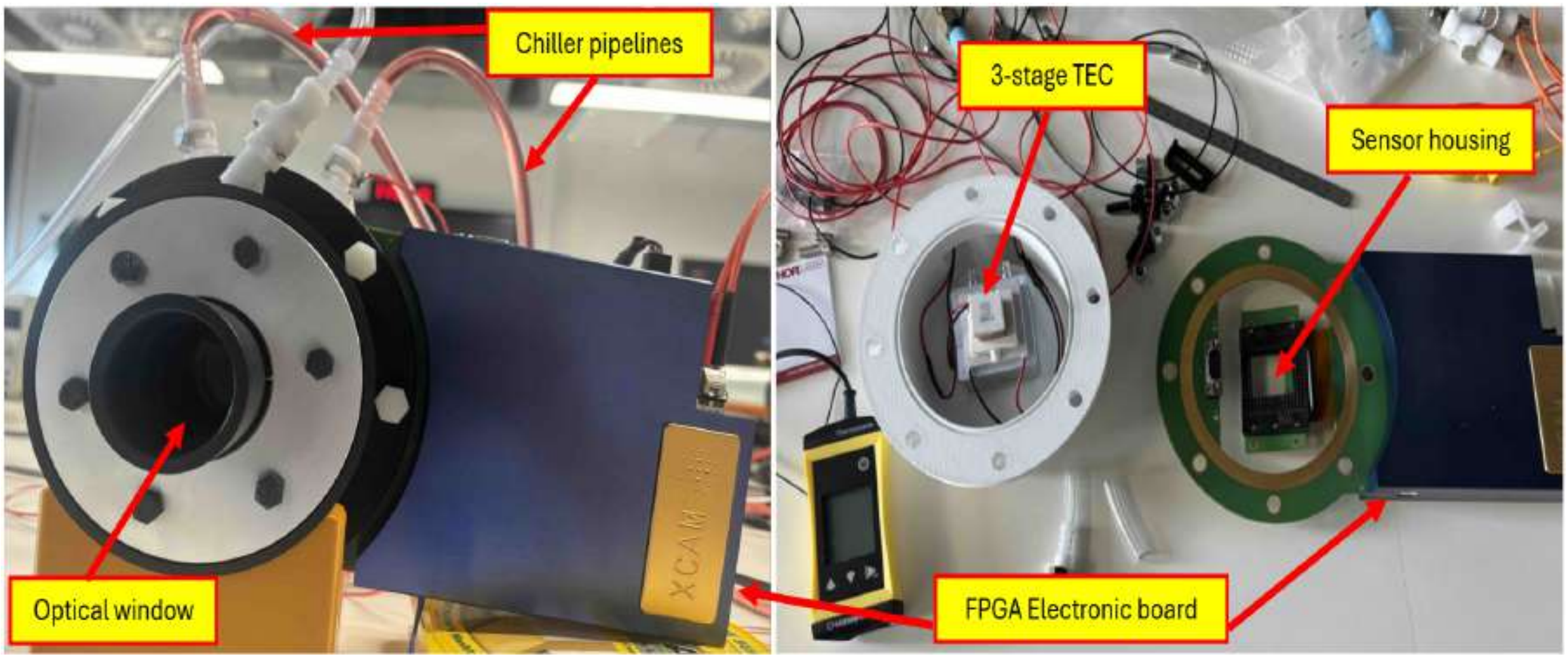

**Figure 2.** Design for the CIS120 telescope housing system. The assembled housing is shown in the left image, while the internal structure, including the construction of the three-stage TEC and the sensor housing is presented in the right image.

## 2 EXPERIMENTAL SETUP

### 2.1 Housing system

The performance of a sensor is significantly influenced by its readout noise. At room temperature, the noise is fairly high due to the dark signal, which adversely affects DR measurements. To reduce noise, a low temperature-controlled environment is necessary. We implemented two distinct cooled systems to house the sensor: one was for laboratory experiment and the other was a 3D-printed prototype, designed specifically for on-sky observations. The desired temperature was achieved using a combination of thermoelectric coolers (TECs) operated with constant current, and a commercial liquid chiller.

#### *2.1.1 Laboratory housing system*

The laboratory housing system was a custom stainless steel casing to hold the CIS120 board, manufactured by XCAM Scientific and provide a stable vaccuum system interface. In order to provide cooling, a two one-stage TECs was used inside the housing. The first temperature sensor was embedded with the CMOS sensor housing, which was mounted directly onto one TEC (referred to as TEC2). The TEC2 was placed on one side of a copper heat sink. The second temperature sensor and the other TEC (referred to as TEC1) were placed on the opposite side of the heat sink. The whole assembly was housed inside a vacuum chamber to minimize condensation and was connected to a liquid chiller. Thorlabs coolflow DTX coolant was used in the chiller for efficient cooling. The front and back sides of the housing system are shown in Fig. 1. However, it was not feasible to attach this housing system to the observatory for on-sky investigations. Thus, we constructed a separate, 3D-printed version with a compatible design for telescope use.

#### *2.1.2 Telescope housing system*

Our 3D-printed telescope housing system employed a three-stage TEC, incorporating a two-stage TEC in place of the one-stage TEC2

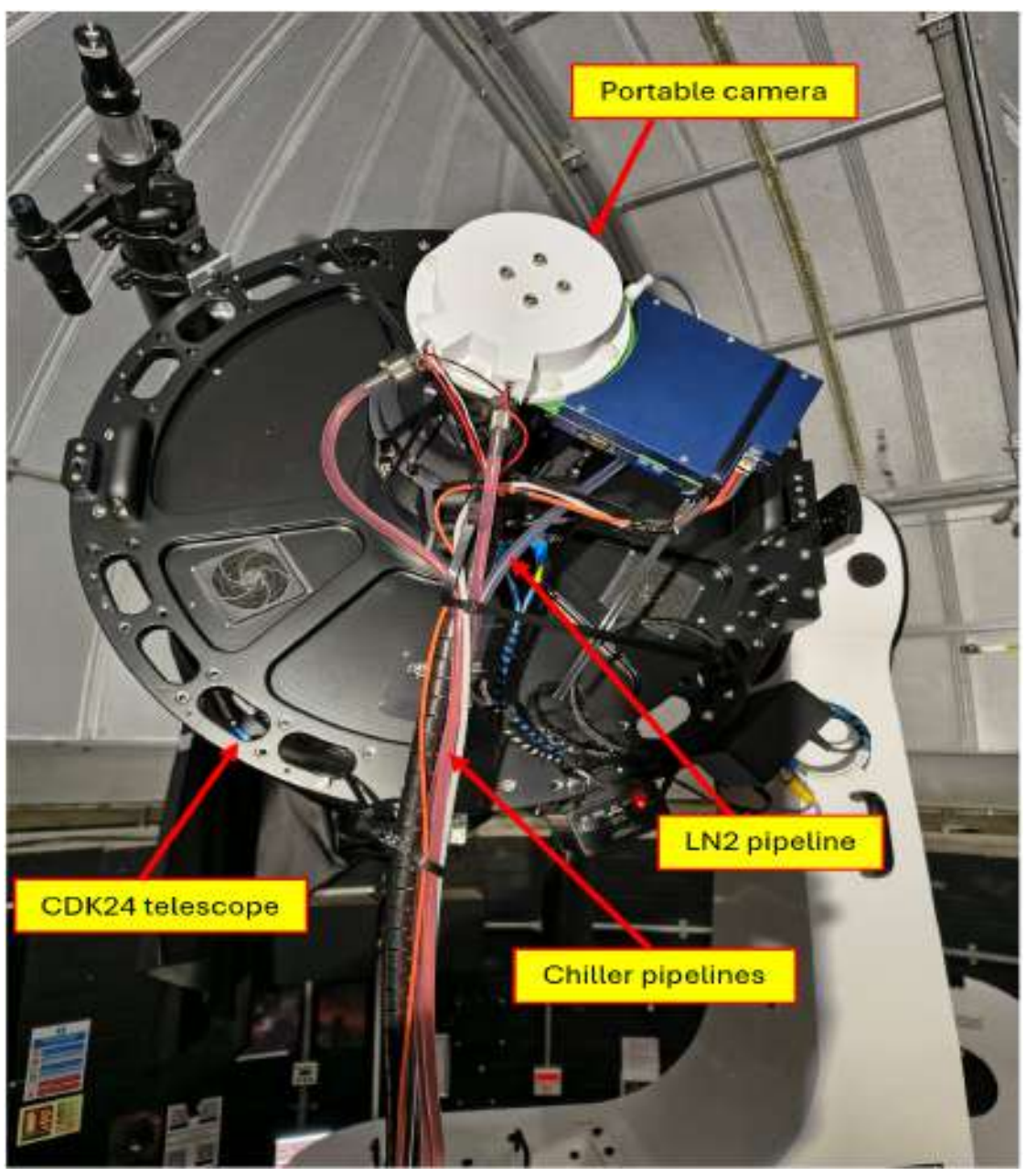

**Figure 3.** CIS120 based 3D printed camera mounted on the CDK24 telescope at Bayfordbury observatory for on-sky testing.

used in the laboratory setup to enhance cooling, while retaining the same one-stage TEC for TEC1. In addition, TEC1 was mounted on top of a Thorlabs cooling block, and the other side of the block was connected to the chiller. To avoid condensation, we opted for circulating dry nitrogen ($N_2$) rather than using a vacuum pump. The telescope housing is depicted in Fig. 2. For on-sky investigations, we mounted the camera on the 60-cm Planewave CDK24 telescope at Bayfordbury Observatory, operated by the University of Hertfordshire, as shown in Fig. 3.

### 2.2 Software

The sensor was mounted on a CIS120 Mk6 PCB, manufactured by XCAM Scientific and controlled via a PCIe-1427 frame grabber card. A Python-based software was developed for establishing communication between the board and the PC. The arbitrary row address was set using a customizable address lookup table, which could be extended up to 10K addresses in size. Our complex readout scheme, as detailed in Section 4.1, was executed by updating the row address for each readout.

The readout operated in two modes: complex mode and standard mode, allowing users to customize and run in the respective mode. Each mode offered several configurable variables to control the operation. Two different lookup tables were used for the two modes. The flow diagram of the software to accumulate experimental data in the complex mode is presented in Fig. 4. The variable 'mini_frame_loops' ($N$) controlled the number of readouts for the bright region, while 'mini_frame_rows' ($R_B$) set the number of rows for the bright region. The variable 'final_frame_loops' ($R_F$) set the number of rows for the faint region or for the entire 2K × 2K frame.

To analyze our acquired data, specifically for DR enhancement experiments, we developed a semi-automated Python notebook[2]. This script applies all necessary corrections, for example, bias, dark and scatter correction and then separates out readout of bright and faint regions, as detailed in Section 4.1. For the bright region, individual readouts of the bright rows are sliced, stacked, and summed, and a cut through the peak of the summed counts is extracted. Similarly, for the faint region, a cut is extracted through the peak of the third transmitted image (see Section 4.2.1 for additional details).

[2] https://github.com/astrosupriyo/pyROWSTACK

### 2.3 Optical setup for in-lab testing

To test the feasibility of our DR enhancement scheme (as explained in Section 4.1), we emulated a scenario involving both bright and faint sources using the experimental setup illustrated in Fig. 5. In this configuration, a collimated 635 nm, 1 mW (max) laser (Thorlabs PL202) was used as the light source. A pinhole wheel (Thorlabs PHW16) and a 50.8 mm diameter super-polished N-BK7 parallel plate beamsplitter (Thorlabs part number WL12012-SP) were implemented to produce images across the sensor. To control the intensity of light illuminating the sensor, a neutral density (ND) filter (Thorlabs NE20A) and a variable ND filter (Thorlabs NDC-50C-4M) were placed in the optical path. The beamsplitter was positioned at a small angle relative to the optical axis to minimize unwanted back reflections from the camera and internal scattering within the camera housing.

## 3 CIS120 CHARACTERIZATION

### 3.1 Bias

We followed Alarcon et al. (2023) to test the stability of bias frames. The spatial and temporal behaviors of each pixel were investigated using 100 bias frames acquired at a sensor temperature of −10°C and at a low-gain setting. For each pixel, the mean and standard deviation from all frames were measured and are shown in Fig. 6. Both distributions are expected to be Gaussian. However, a slight deviation from the ideal Gaussian shape was apparent. Variations in the mean indicate different bias voltages, while the spread in standard deviation reflects read noise differences across pixels.

To generate a master bias, all 100 bias frames were median combined. A master bias is presented in Fig. 7 for an example. The resulting master bias has a mean of 221 ADU or 2652 $e^-$ and a standard deviation of 4.3 ADU or 51.6 $e^-$. Pixel-to-pixel count variation was evident in the master bias, arising from the different readout and thermal noise characteristics of the sensor's built-in electronics in each pixel. Therefore, we used a full array of bias or dark frames for subtraction in subsequent analyses, ensuring accurate correction of these non-uniformities.

### 3.2 Read noise

We acquired a total of 100 consecutive bias frames in low-gain setting to determine the sensor's read noise. Data acquisition was performed at −10°C. By subtracting each pair of consecutive frames, we generated 99 difference frames, effectively removing the bias level from each image. For each pixel, we then computed the standard deviation from all 99 frames. The resulting distribution of standard deviations was divided by the square root of two to obtain the read noise distribution. Fig. 8 shows the histogram of the read noise. The mean (median) read noise was measured as 0.85 ADU or 10.2 $e^-$ (0.84 ADU or 10.1 $e^-$), respectively.

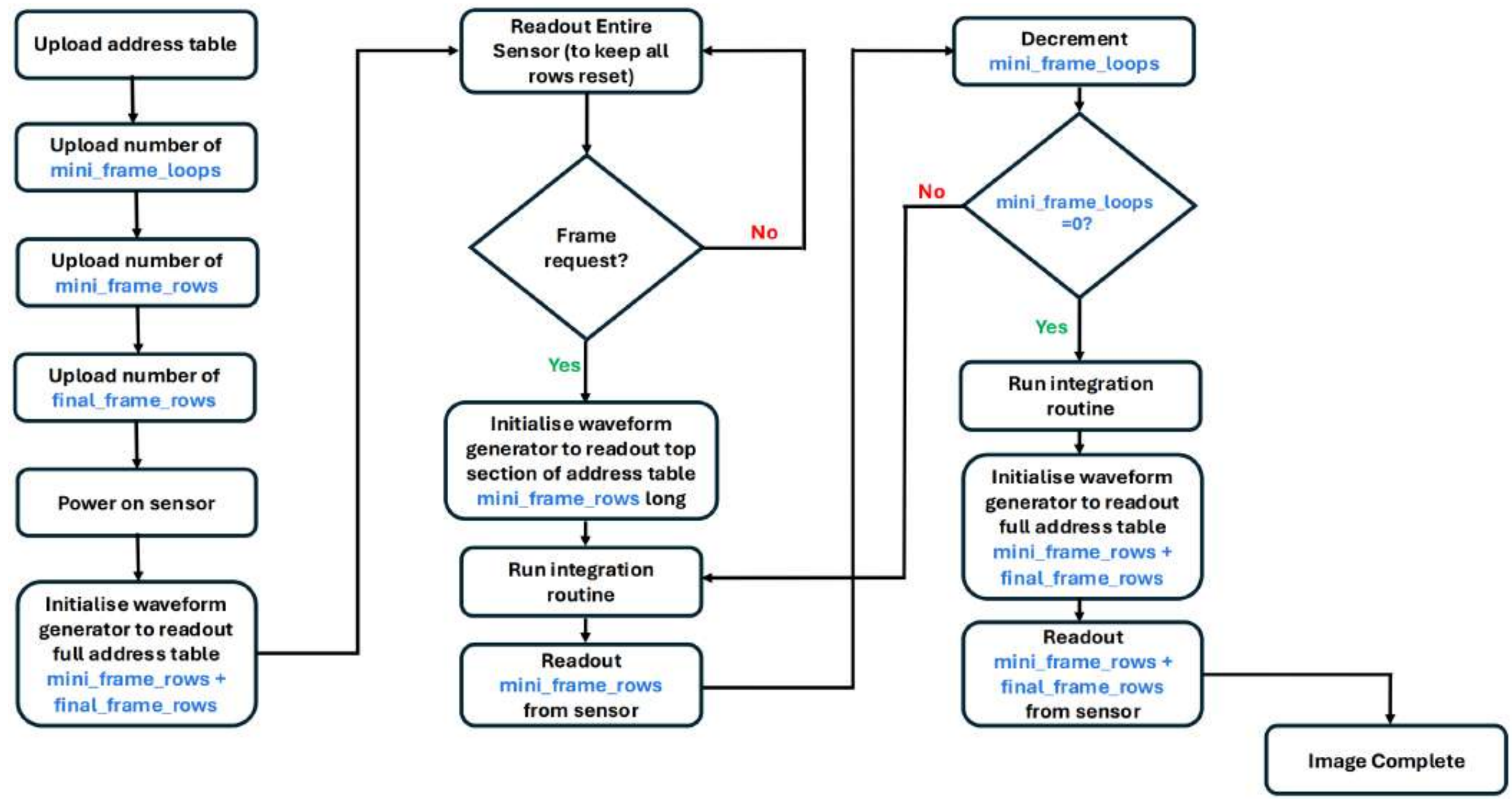

**Figure 4.** Sequencing of the detector for complex read-out mode.

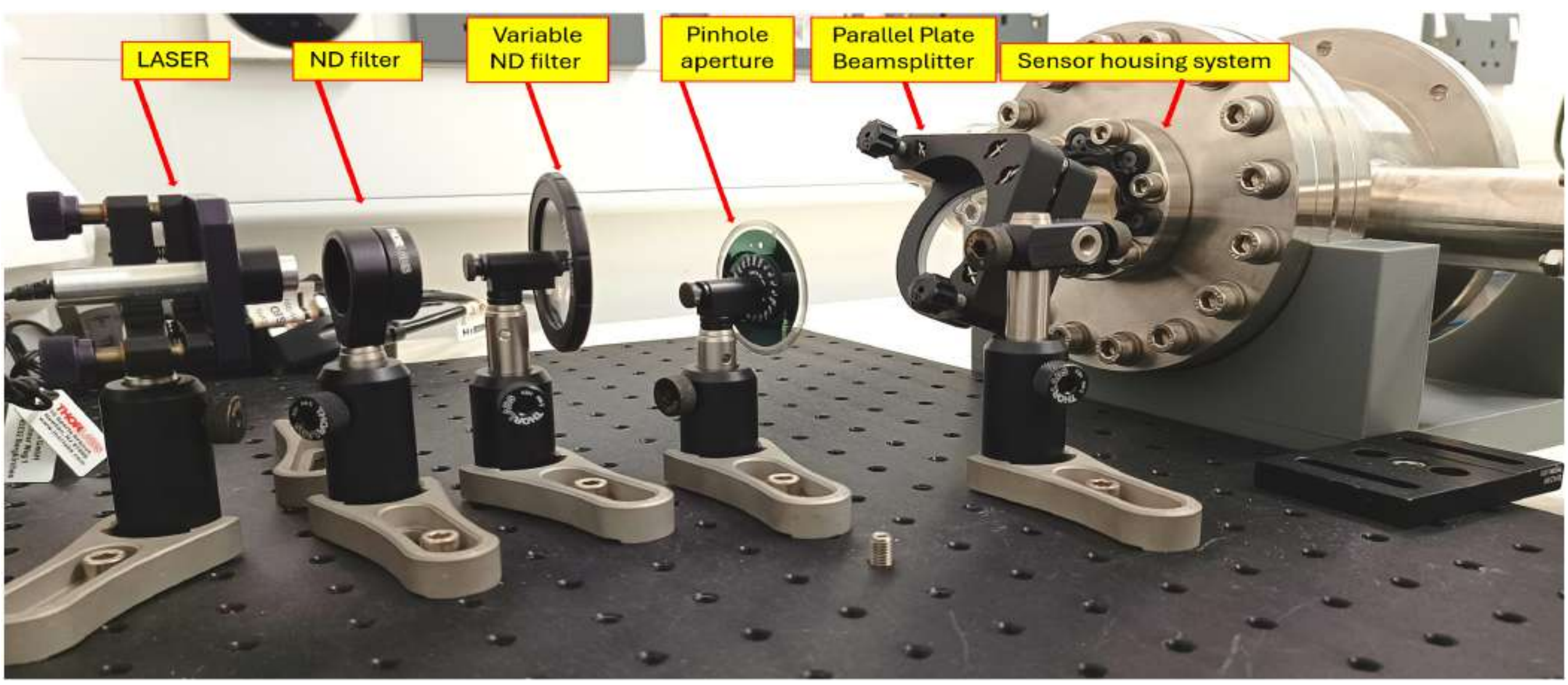

**Figure 5.** Optical setup to form transmitted images on the sensor.

### 3.3 Dark current

Several dark frames were captured in the laboratory at varying integration times from 0 ms to 2000 ms in high-gain setting. In each integration time, multiple dark frames were average combined to form a master dark, and then bias-subtracted. The mean count of each bias-subtracted frame was calculated from a 700 × 700 pixel window and plotted as a function of integration time for each temperature. A straight line was fitted to these data points, and the slope of the fit provided an estimate of the dark current at a certain temperature. This procedure was repeated across temperatures from +20°C to −50°C. The resulting plot of dark current versus temperature is shown in Fig 9. At −50°C, the mean dark current reached a value of 0.004 $e^-$/s.

### 3.4 Linearity

To obtain data for linearity correction, a halogen broadband lamp was used to provide uniform illumination to the sensor. The data acquisition was carried out in the laboratory at −30°C with the integration time from 0 s to 14 s in steps of 0.5 s in high-gain setting. For each integration, a 700 × 700 pixel window was selected to calculate the mean count. To correct for bias and ensure the linearity curve passed through the origin, the mean count of the first frame (obtained at zero second exposure) was subtracted from all frames, including the zero-second frame itself.

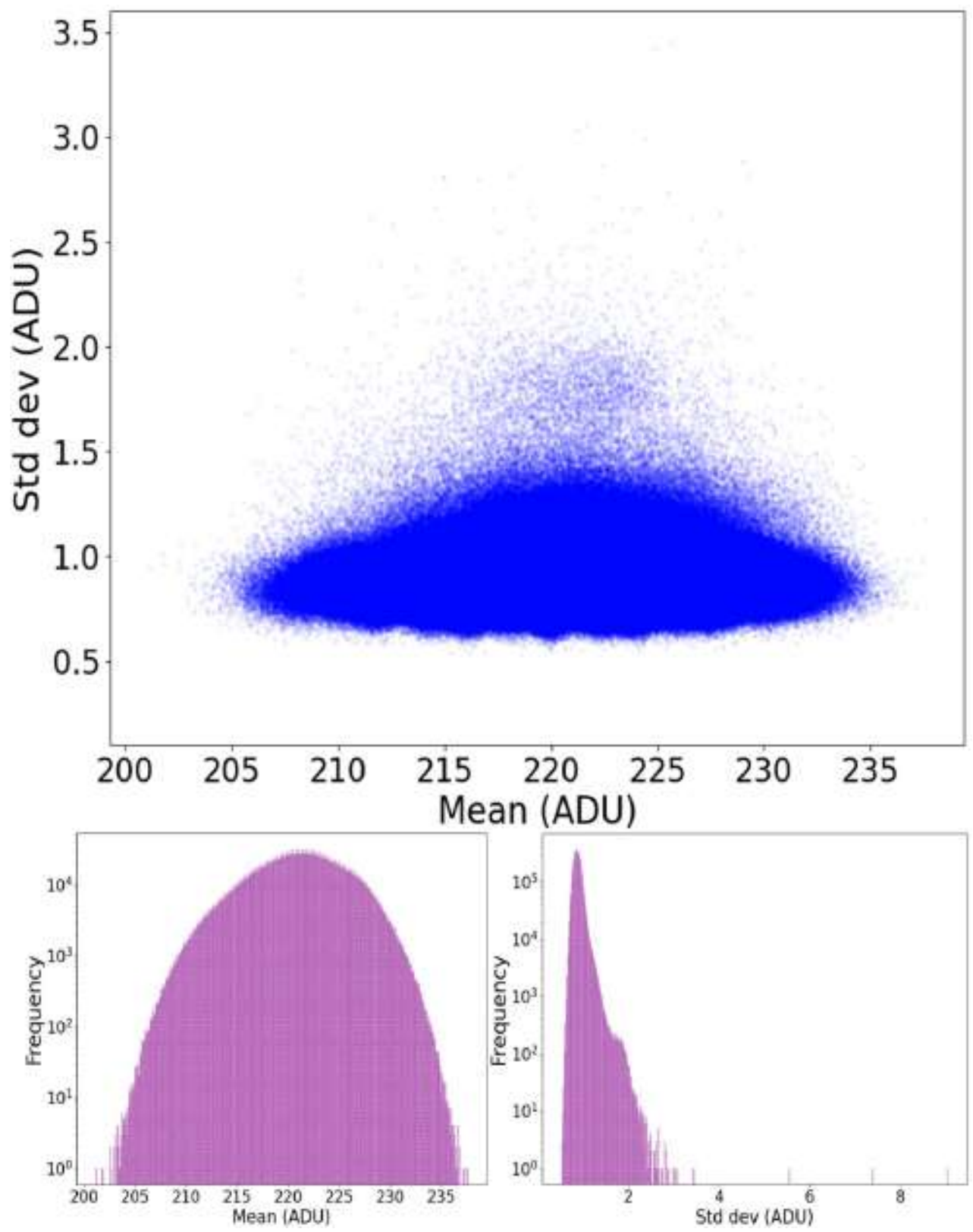

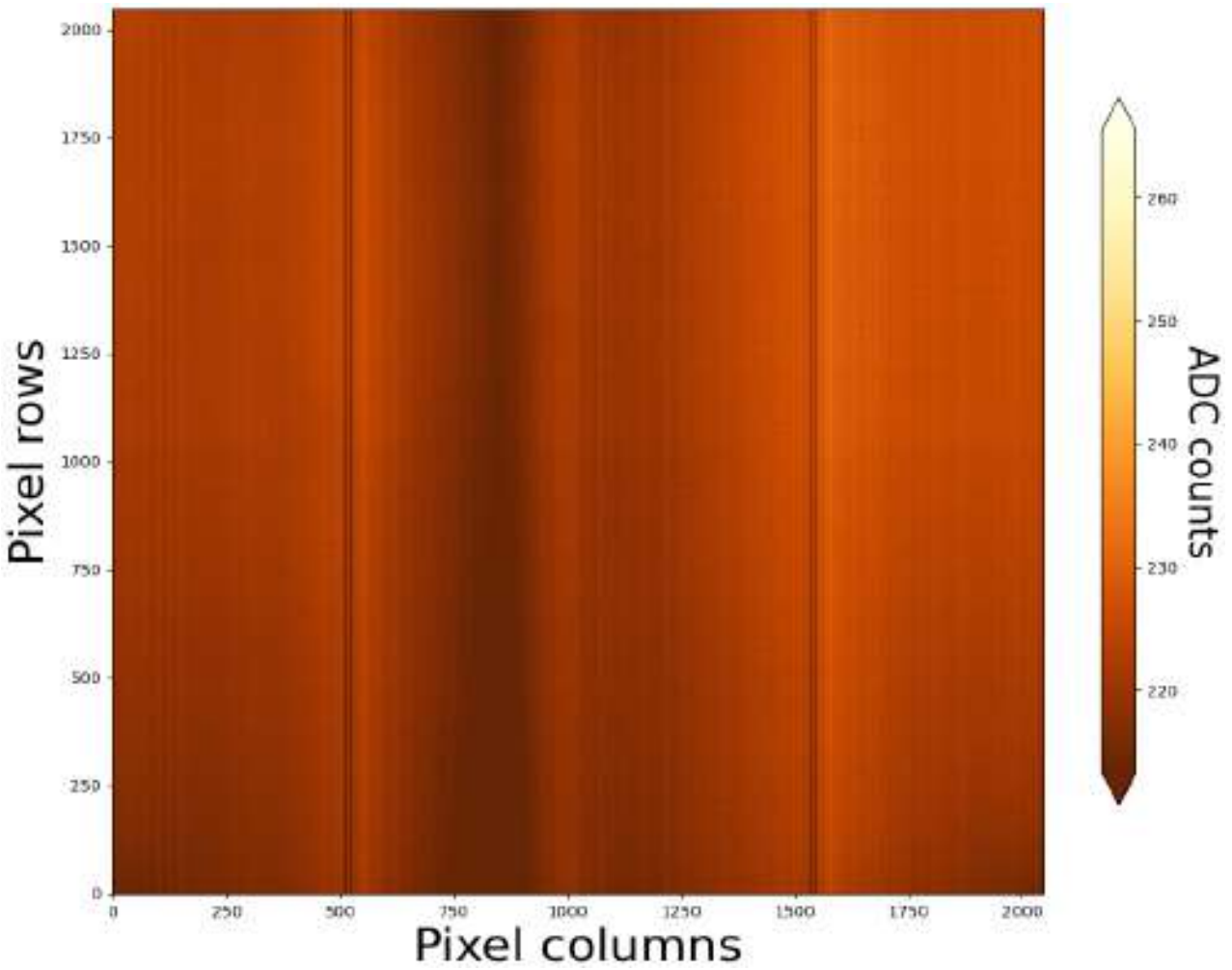

**Figure 6.** Top panel: mean versus standard deviation of each pixel for 100 consecutive bias frames to investigate the temporal performance of the CIS120 (top panel). Bottom panel: histograms of pixel-wise mean (left) and standard deviation (right) distribution. The y-axis of both histograms is in log scale.

**Figure 7.** Master bias of 100 consecutive bias frames (median stack) is presented to show the spatial behavior of each pixel. Pixel-to-pixel count variations are apparent.

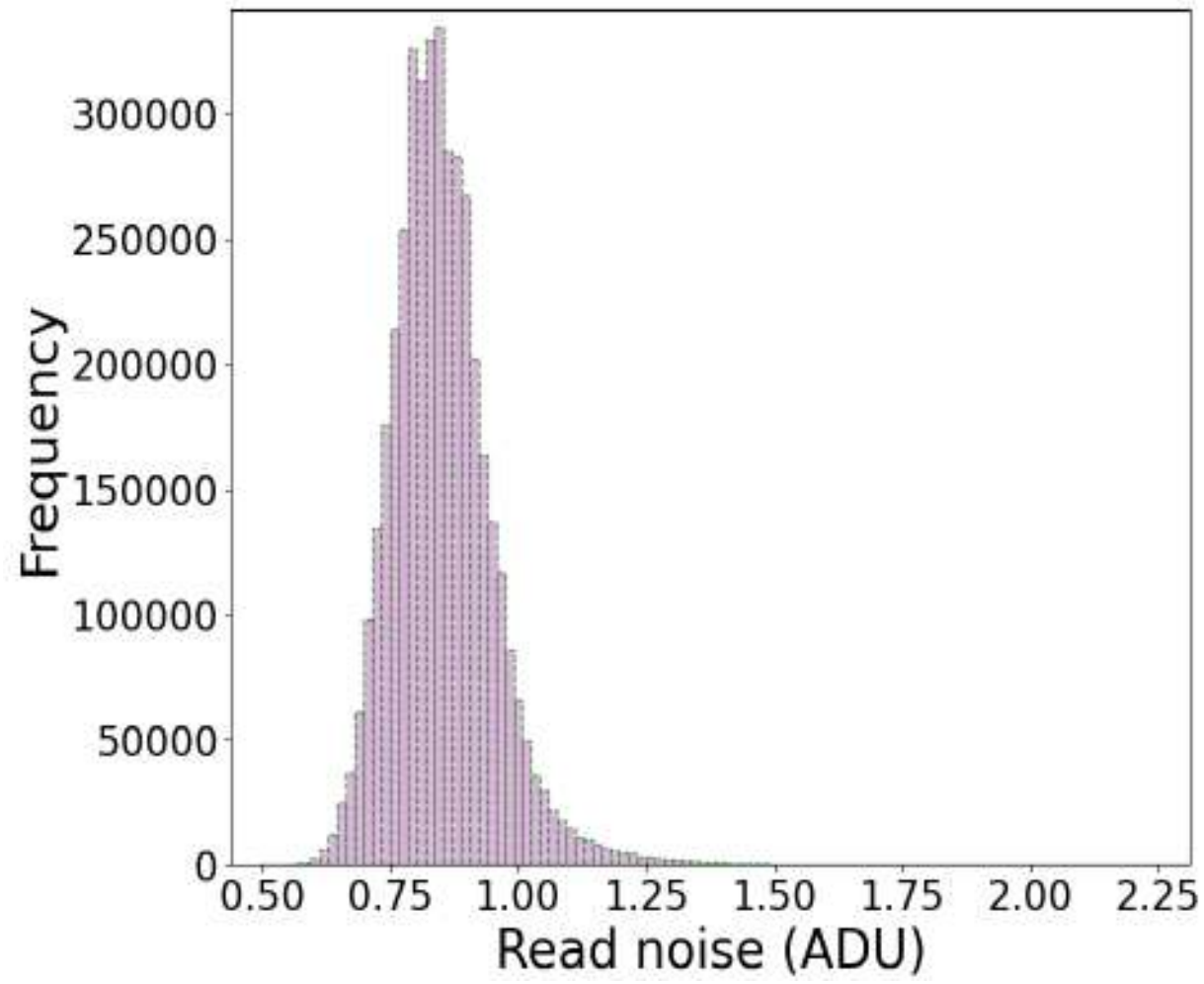

**Figure 8.** Read noise measurement. The median and mean read noise are 0.84 ADU and 0.85 ADU or 10.1 e$^-$ and 10.2 e$^-$, respectively.

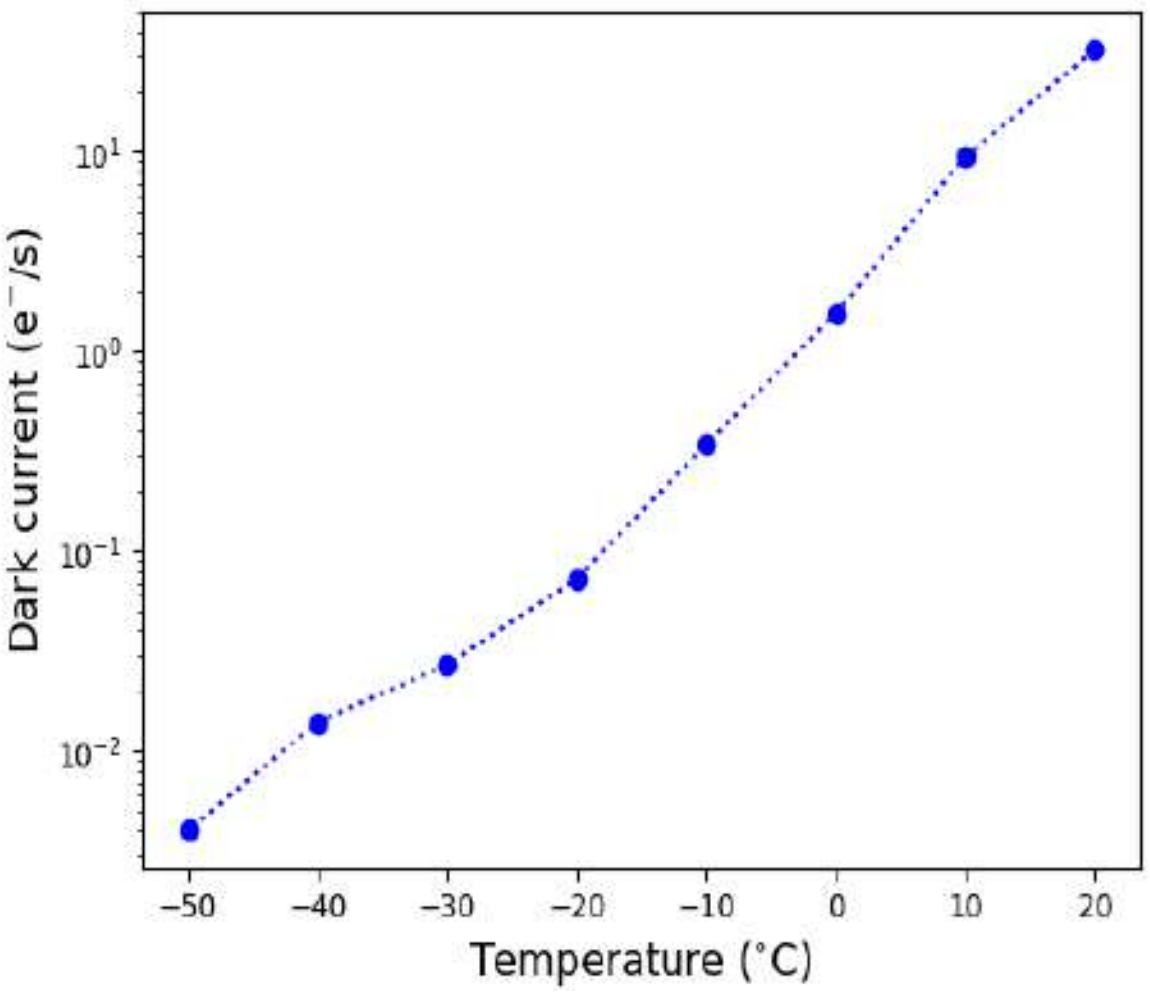

**Figure 9.** Measured mean dark currents at different temperatures of CIS120.

The difference in counts between consecutive frames was calculated, from which the difference in measured count rate was derived. The left panel of Fig. 10 illustrates the differential counts versus measured mean counts. Notably, the initial five measurements display a constant count rate. Using these initial readouts, a linear model was fitted to establish the linearity correction. This correction is shown by the black line in the middle panel of Fig. 10 and the corresponding residuals is displayed in the right panel. We note that the last five measurements deviate by more than 20% from the initial non-linearity correction curve, showing a significant deviation from linear behavior. Furthermore, the normalized difference in measured count rate relative to in-frame counts shows a change of less than 1%, indicating the beginning of detector saturation. Thus, we excluded these points from further analysis.

Various models were previously implemented to correct detector non-linearity (see, e.g., Ninan et al. 2018; Petric et al. 2024). In this work, we followed the approaches described in Soman et al. (2015) and Wocial et al. (2022) for non-linearity correction. Rather than applying a quadratic fit, a cubic fit was used to more accurately model the data. The inverse of this cubic function was then used to convert the ADU to the time domain, which is proportional to illuminance. The cubic fit is presented by the green line in Fig. 10. The final non-linearity correction is depicted by the red line. Notably, the linearity correction derived from the initial few readouts is consistent with the final linearity curve.

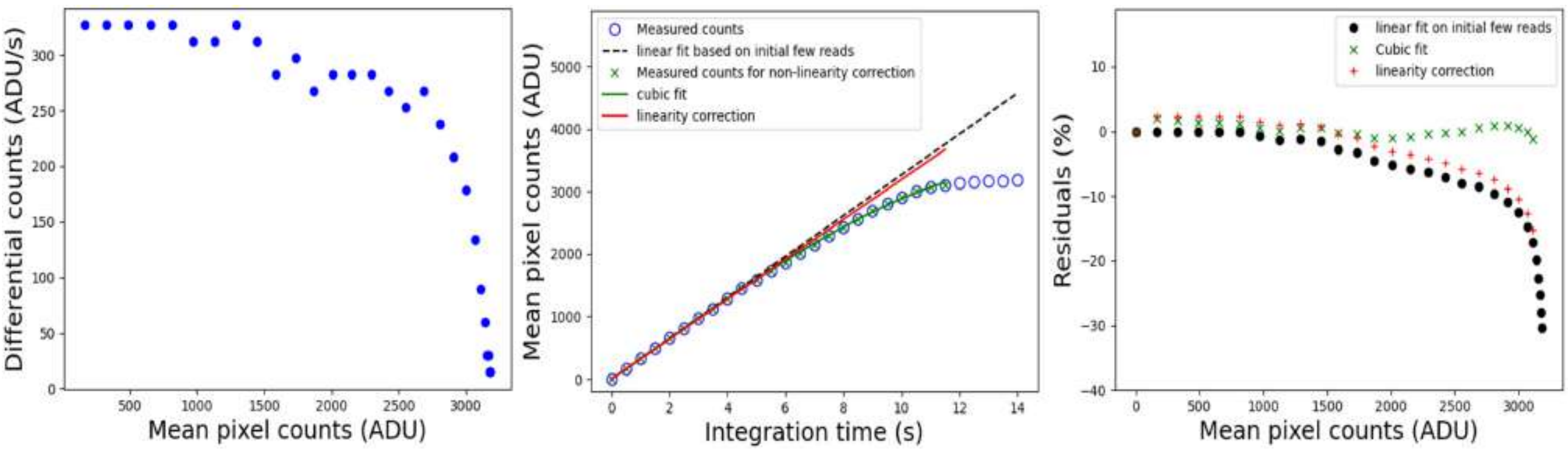

**Figure 10.** Linearity analysis of CIS120 is shown. The left panel displays the difference in measured count rate versus measured counts. The middle panel depicts the mean counts within a 700 × 700 pixel window of a uniformly illuminated region after bias subtraction plotted against integration time. The black line represents the initial non-linearity correction derived by fitting a linear model to the first six readouts (including (0,0)). The last five reads were excluded from further analysis because of a small increment in difference in measured count rate and a large deviation (greater than 20%) from the initial non-linearity correction curve. A cubic model was fit for non-linearity correction, indicated by the green line in the plot. The post-corrected linearity curve is illustrated by the red line. The right panel displays the residuals for all three curves described earlier for a visual inspection of the linearity analysis.

### 3.5 System gain

System gain was determined using the single-electron detection technique described in Soman et al. (2015). X-ray events from a radioactive $^{55}$Fe source at −50°C were recorded in the laboratory in a high-gain setting, as shown in Fig. 11. The decay of $^{55}$Fe to $^{54}$Mn produced X-ray photons, which generated a total of 1620 e$^-$. Two prominent peaks are visible at 1.81 ADU (noise peak) and 917.89 ADU ($^{54}$Mn $K_\alpha$). Based on these measurements, the system gain was calculated to be 1.77 e$^-$/ADU. The low gain setting is measured in a similar way, the system gain was 12.0 e$^-$/ADU.

### 3.6 Dynamic range

As shown in Fig. 10, after applying linearity correction, the maximum achievable signal is approximately 3670 ADU or 44040 e$^-$ in the low-gain setting. Beyond this point, the detector appears to be saturated. Therefore, the maximum dynamic range attainable with this 12-bit sensor in standard readout mode is approximately 71 dB. In the following section, we present how this native dynamic range can be significantly enhanced.

## 4 ENHANCING DYNAMIC RANGE

### 4.1 Techniques

In the complex mode readout, sensor rows were divided into two groups, each with different effective integration times. An arbitrary number of rows ($R_B$ – bright rows) can be selectively addressed, read out, and reset multiple times ($N$) with a chosen integration time ($t$), without interrupting the ongoing photon collection in the remaining rows ($R_F$ – faint rows). The total number of photons collected by $R_F$ increases as its integration time ($T = t \times N$) increases, in proportion to both $t$ and $N$.

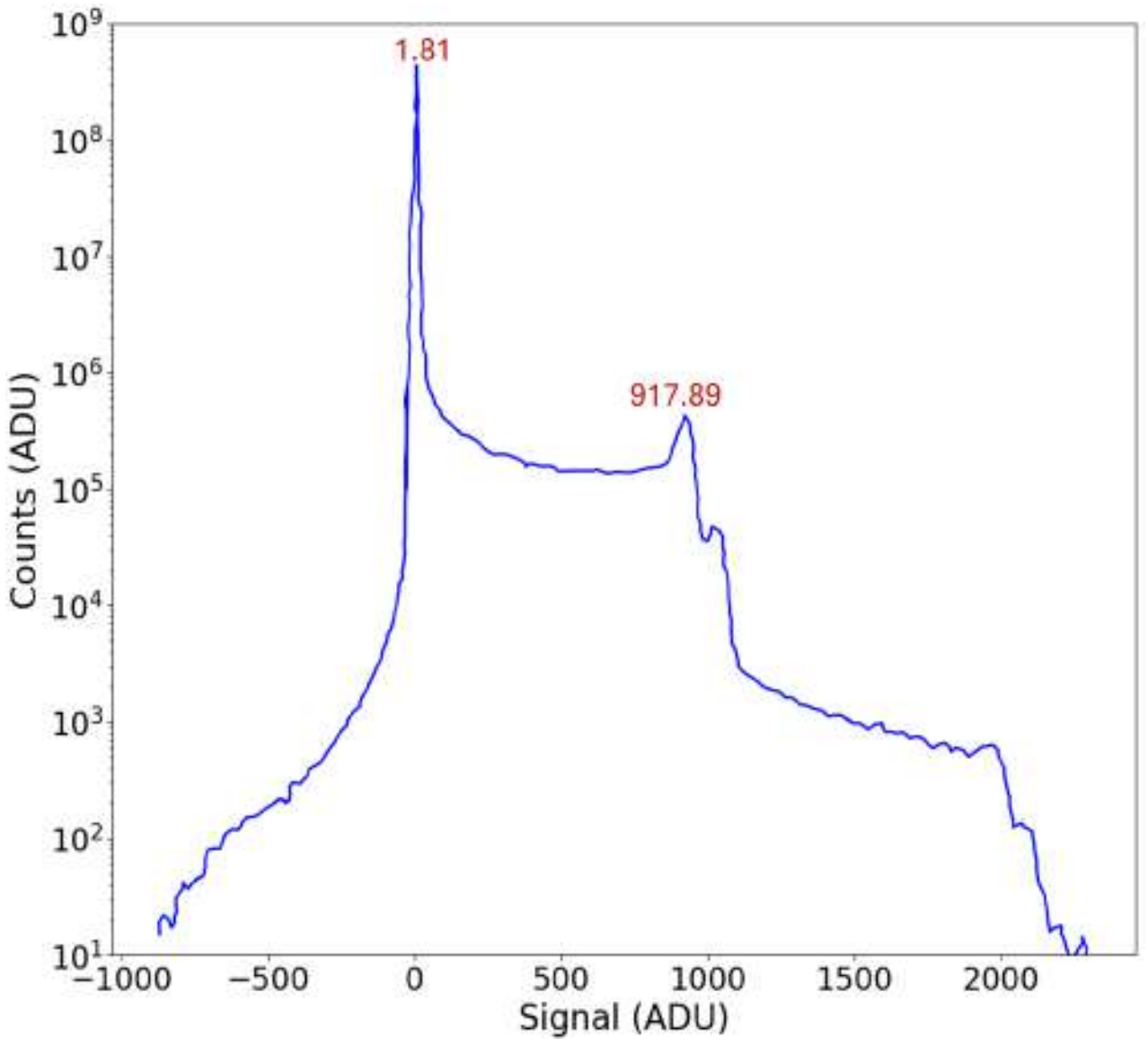

**Figure 11.** Gain measured using X-ray events from a radioactive $^{55}$Fe source is shown at −50°C. The plot displays two marked peaks: 1.81 ADU, corresponding to the noise peak, and 917.89 ADU, corresponding to the $^{54}$Mn $K_\alpha$ peak.

This scheme allows longer exposures for faint sources while preventing saturation of bright sources, thereby extending the sensor's DR by a factor of $20\log(N)$ dB. The FWC for $R_B$ effectively increases by a factor of $N$, although the read noise in $R_B$ also increases by $\sqrt{N+1}$. Additionally, as $R_B$ is read out multiple times, the total read-out time and the overall integration time increase by a factor of $N$. Additional details can be found in Wocial et al. (2022). An example of this multiple-row read-out scheme is illustrated in Fig. 12 for $N = 14$.

### 4.2 Results

#### *4.2.1 Laboratory experiment*

We carried out our in-lab test using the laboratory housing system and optical setup as described in Sections 2.1.1 and 2.3, respectively.

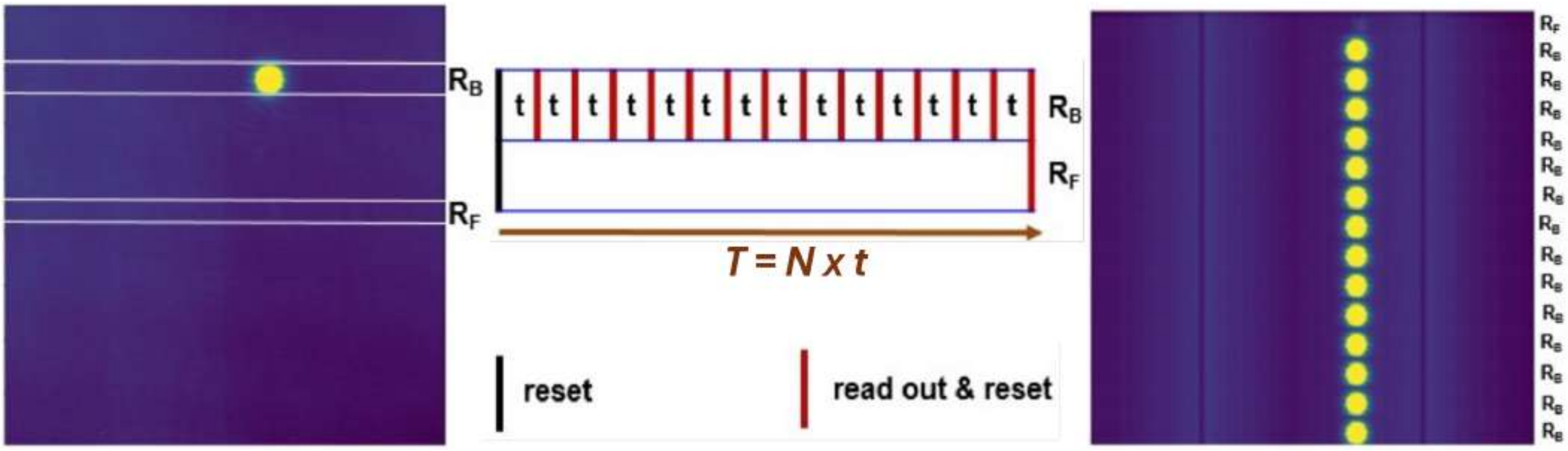

**Figure 12.** Illustration of the complex read-out scheme. The left-hand image highlights the bright rows, marked as $R_B$, where a bright spot was formed by the first transmitted image. These rows were read out and reset multiple times (here, $N = 14$ as an example), while the faintest region, marked as $R_F$, continued to accumulate photons, as shown in the middle image. This was then followed by the read-out and reset of the faintest region. Finally, in the right-hand image, the final image produced by the complex mode is illustrated: the faint spot, $R_F$, is barely visible above 14 times bright rows read-out.

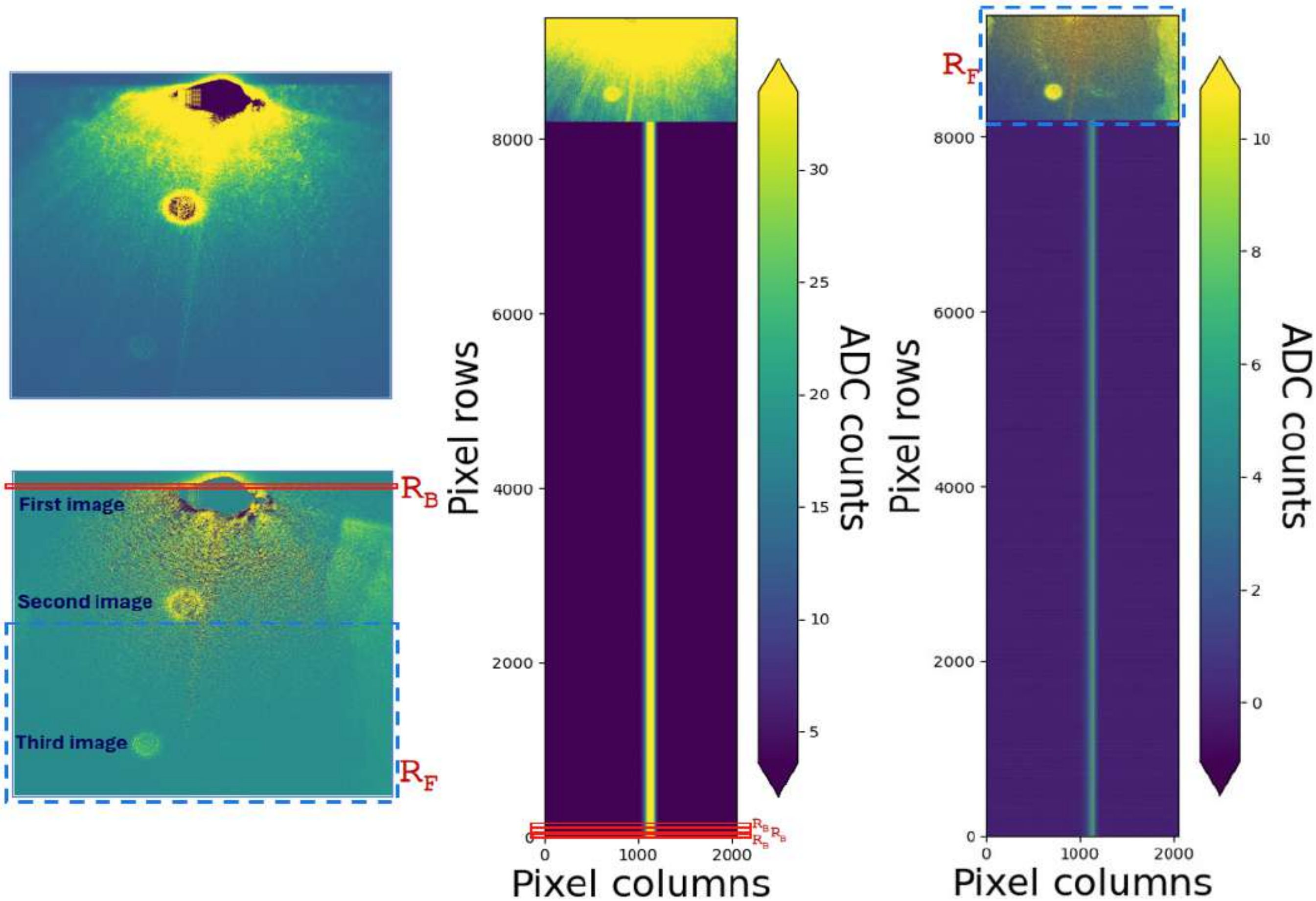

**Figure 13.** Demonstration of the data acquisition and analysis method for the laboratory experiment. First, an image is obtained in the standard mode (top-left). This image is then bias, dark, and scatter-frame subtracted (bottom-left) to prominently identify the first and third spots for running the software in the complex mode. The three transmitted images, obtained using the optical setup illustrated in Fig. 5, are marked. The first and third spots are designated as the bright ($R_B$) and faint rows ($R_F$), respectively, for data acquisition in the complex mode. The middle image shows the obtained bias- and dark-subtracted raw frame in complex mode; all $R_B$ readouts (three are marked as examples) from this frame are sliced and summed to extract a spectrum of a bright row, as shown in Fig. 14. The right image displays the bias-, dark-, and scatter-frame subtracted raw frame in complex mode, from which the marked $R_F$ was used to extract a spectrum of a faint row, also shown in Fig. 14.

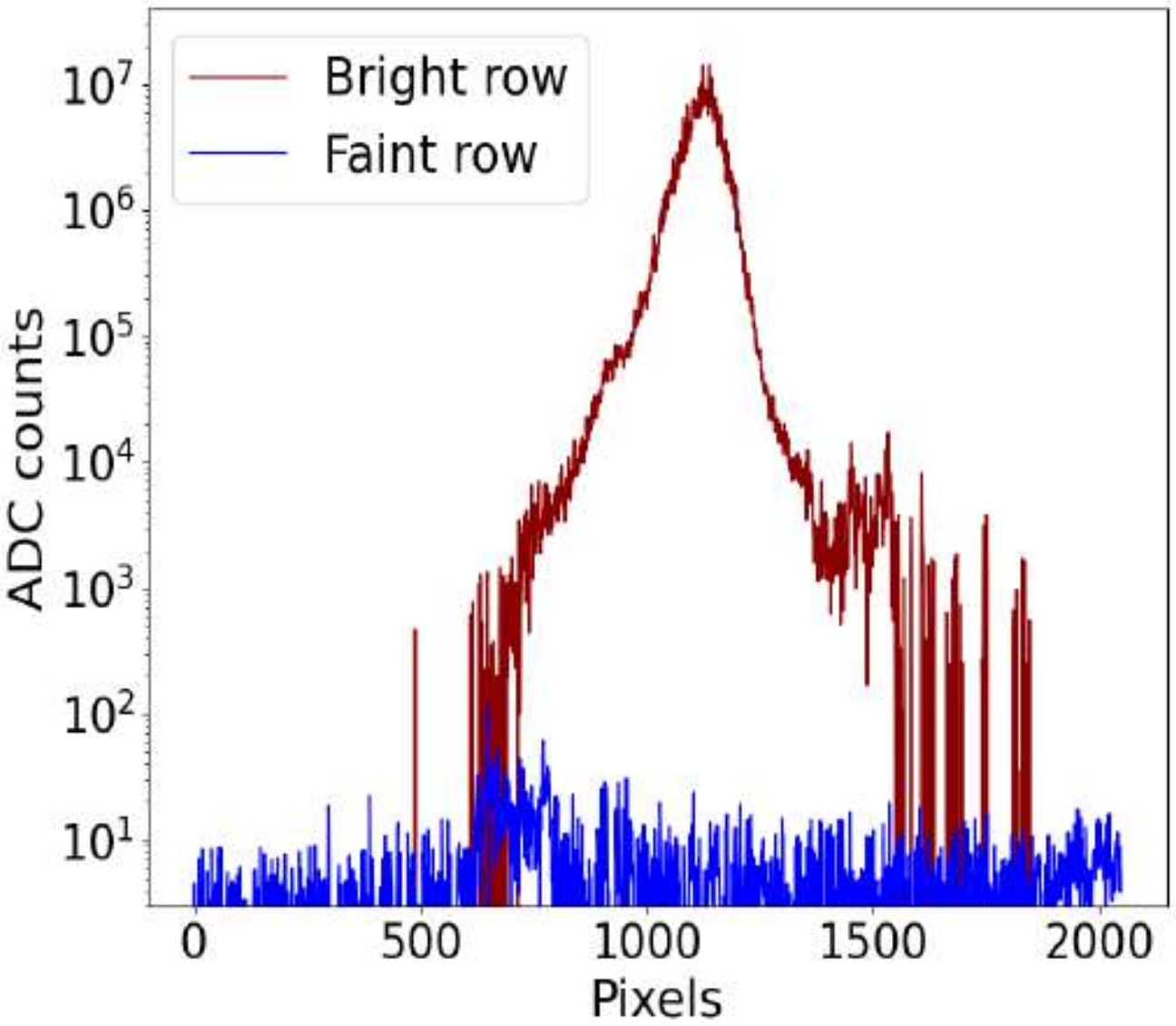

**Figure 14.** Demonstration of the complex read-out scheme in the laboratory experiment. The brown line represents a cut through the peak of the summed counts of $N$ $R_B$ slices, generated via multiple read-outs as illustrated in Fig. 13. The blue line shows a cut through the peak of the third transmitted image, which was read out once, as indicated by $R_F$ in Fig. 13. It is evident that the difference in counts between the first and third peaks is approximately $10^7$.

In general, the beamsplitter used in our optical setup has a reflectance of less than 4%. Assuming the maximum reflectance of 4%, the intensity ratio between the first transmitted beam and that of the second beam is approximately $10^3$:1, while the ratio between the first and third transmitted beams is about $10^7$:1. In our experiment, we aimed to probe the first and third images as bight and faint spots, respectively. However, the large intensity difference between these images, combined with background noise and scattering light, made it challenging to detect both simultaneously on the detector during a single exposure. To minimize sensor background noise, we carried out the experiment at $-10°$C. Additionally, we optimized the alignment of the experimental setup and adjusted the incoming light intensity and integration time until both the first and third spots became clearly visible.

The experiment was performed in a low-gain setting because low gain provides a wider dynamic range and a higher full-well capacity. Initially, data were acquired in the standard read-out mode for a certain exposure, but the level of scattered light was so high that the third spot was hardly detectable. To mitigate this, we employed two strategies: (i) optimizing the alignment of the optical components so that part of the central spot from the first beam was positioned at the edge of the sensor, and (ii) introducing an aperture mask between the beamsplitter and the sensor housing to selectively block the path of the third beam. Data were obtained for same exposures both with and without the aperture mask, referred to as the scatter frame and object frame, respectively. Additionally, we acquired multiple bias frames and corresponding dark frames. After correcting the bias and dark for both object and scatter frames, we subtracted the scatter frame from the object frame to detect the third spot. The raw and processed object frames obtained in the standard read-out mode are displayed at the top and bottom of the left panel of Fig. 13, respectively.

It is important to note that, when both the first and third spots were clearly detected, the bright spot was highly saturated, while the third spot was just at the threshold of detection (see, the top left panel of Fig. 13). This observation clearly demonstrates the limitation imposed on astronomical observations due to the limited DR of a 12-bit camera in the standard readout mode.

Once the first and third spots were clearly visible, we selected bright and faint regions to acquire data in the complex read-out mode. The bright rows ($R_B$), each of $2 \times 2$K, were read out a total of $N = 4097$ times, while the faint rows ($R_F$), also $1200 \times 2$K in size, were read out once. For illustration, $R_B$ and $R_F$ are indicated by red and blue boxes, respectively, in Fig. 13. The final image size was $9394 \times 2$K. Object, scattered, bias, and dark frames were obtained in the complex read-out mode using the same procedure described above for the standard read-out mode.

The object frame was first bias- and dark-subtracted as shown in the middle panel of Fig. 13. Each $R_B$ readout from the object frame was then sliced and stacked, resulting in a total of 4097 slices. These slices were summed on a pixel-by-pixel basis. If the peak count of a single slice is $C$, then after summing $N$ slices, the peak count becomes $N \times C$ (assuming no brightness variation of the light source during readout). No brightness variation is expected in the laboratory experiment; however, this might not be the case during on-sky experiments. Nevertheless, from the summed $R_B$ of $N$ slices, a single row was extracted and plotted in brown in Fig. 14. The maximum peak count ($C_{max}$) was determined to be on the order of $10^7$.

For $R_F$, we subtracted the bias- and dark-corrected scattered frame from the bias- and dark-corrected object frame. A row at the center of the third spot was then extracted and plotted in blue in the left panel of Fig. 14. In this case, the minimum peak count ($C_{min}$) was on the order of $10^1$. The DR, calculated as $20 \times \log(C_{max}/C_{min})$, achieved from this experiment was approximately 120 dB. Furthermore, we carried out the above experiment in a similar way with $R_B = 1$ and $N = 8190$, achieving a DR of 134 dB. However, the total number of bright rows read is finite and depends on the buffer size, which restricts the achievable DR. Nevertheless, in practice, we prefer to oversample the target's PSFs, so we typically read out at least two bright rows.

The DR could potentially be improved by either increasing the count level of the first spot or minimizing the count level of the third spot. However, in this work, we did not fully optimize these count levels, as our primary objective was to demonstrate the capability of our scheme to enhance the DR for CIS120, rather than to maximize it.

#### 4.2.2 *On-sky experiment*

As depicted in Sec 2.1.2, we mounted our CIS120 camera for on-sky data acquisition. Observations were performed using the Baader SLOAN/SDSS y′-filter, which operates in the range 955−1057 nm [3]. This filter was chosen for its ready availability and its suitability for modern CMOS cameras. Furthermore, the observations took place under nearly full moon in clear conditions; operating in the near-infrared was advantageous in this context, as the effect of sky brightness due to moonlight is minimized at these wavelengths.

We selected two bright stars with magnitudes close to zero, Vega and Polaris, for our observations. For each star, we first obtained a 2K × 2K image in the standard read-out mode with an exposure chosen to avoid saturation. During the observation, we estimated the centroid of the object using SAOImageDS9 (Smithsonian Astrophysical Observatory 2000). A strip of 10 pixel-rows, centered at the measured centroid, was then selected for the complex mode row read-out. This choice allows for centroid detection at every readout in complex mode and can be adjusted based on observing conditions and telescope pointing accuracy. Once the number of rows was determined, the number of complex mode read-outs was limited by the sensor's buffer size; in this setup, 781 read-outs were executed, followed by a full 2K × 2K image as the final read-out. According to our convention, the $R_B$ region consisted of 10 pixel-rows (10 × 2K) read out $N = 781$ times, while $R_F$ was 2K × 2K and reads out once. Therefore, the final image size in complex read-out mode for a single exposure was 9858 × 2K ($R_B \times N$ plus $R_F$), as illustrated in Fig. 15.

[3] https://www.baader-planetarium.com/en/baader-sloan-sdss-y-filter-photometric.html

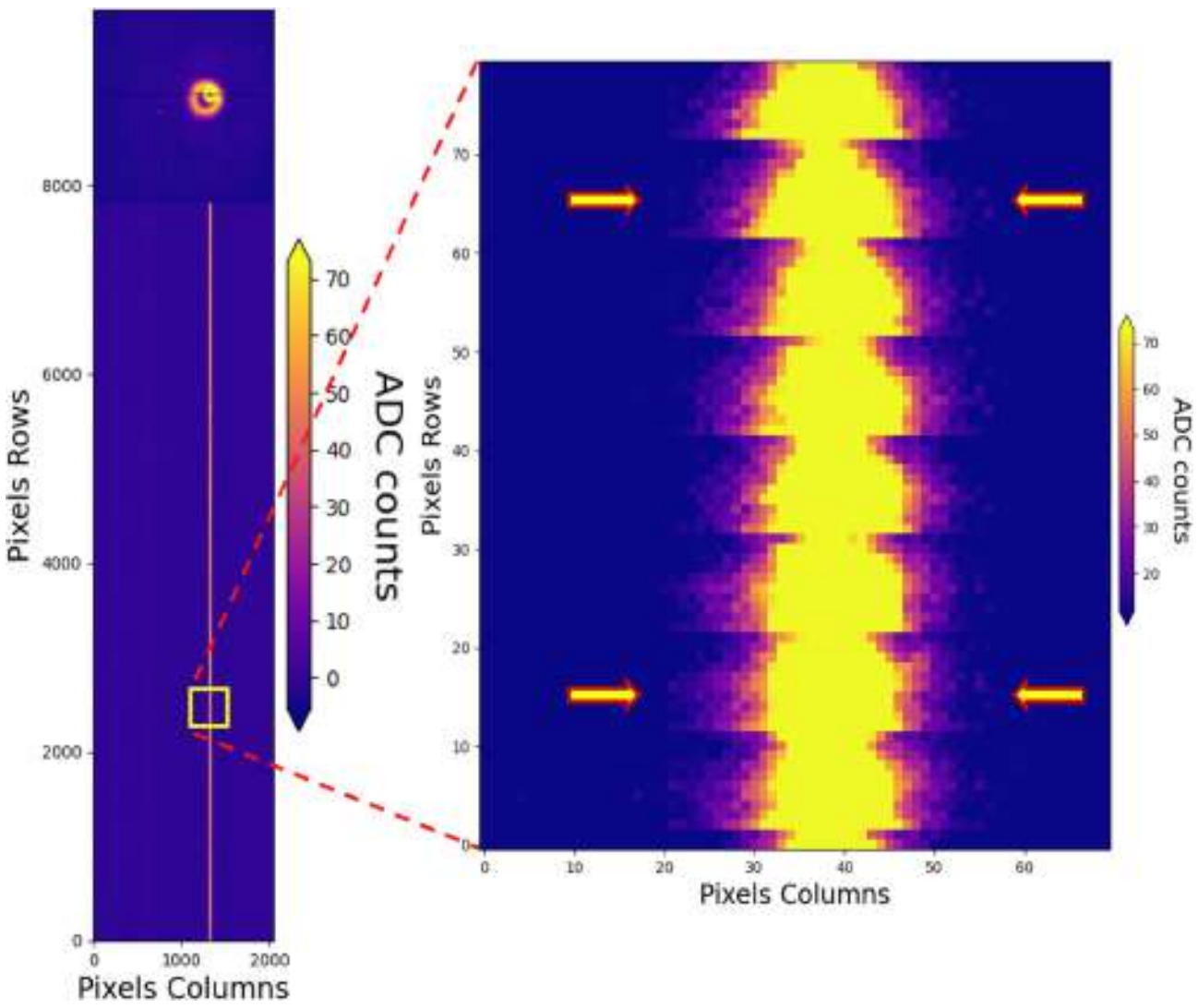

**Figure 15.** The bias- and dark-subtracted full frame observed in the complex mode readout is displayed in the left panel, alongside a zoomed-in portion that demonstrates how the image centroid shifts during fast readout in on-sky observations. The zoomed-in region illustrates the apparent positional changes of the star (here, Vega) during consecutive readouts (right panel). Two specific readouts, indicated by arrows, are shown: the top arrow points to a readout where only half of the PSF is captured within the readout regime, while the bottom arrow indicates a readout where the PSF is almost fully contained within the readout region.

Additionally, we employed a typical five-point dithering method to reduce the near-infrared sky background (see Ferrarese et al. 2012, for details on the dithering process). At each dither position, object frames were acquired in both standard and complex modes as described above. Multiple bias and dark frames were also obtained in both modes for bias and dark subtraction, respectively, during data reduction.

For analysis, the object frames were cleaned by subtracting bias, dark, and sky frames, following the method of Ghosh et al. (2018). The sky frame was constructed by median combining all object frames obtained at different dithered positions. However, the resulting sky frame appeared less uniform than expected, which can be attributed to significant internal scattering within the camera during long effective integration times in the presence of a very bright star (see Section 5 for details). When subtracting the sky frame from the object frames, any negative counts in the sky frame would increase the background level in the resulting images. To minimize this effect, all negative counts in the sky frame were set to zero prior to subtraction.

An example of a cleaned image, observed in standard mode read-out, is shown in the middle panel of Fig. 16 (Fig. 17) for the Vega (Polaris) field, respectively. In the complex mode, all $R_B$ rows in a cleaned object frame were sliced and stacked together. These 781 slices were then summed on a pixel-by-pixel basis and replaced with the corresponding 10 pixel-rows from $R_F$ region. The reconstructed image is displayed in the right panel of Fig. 16 (Fig. 17).

Typical differences in image contrast scenarios between a 16-bit camera and a 12-bit camera are illustrated in Fig. 16 and Fig. 17 for Vega and Polaris, respectively. The observations with the 16-bit camera were acquired in the $R$-filter with exposures of 5 seconds and 60 seconds, respectively, using a ZWO ASI6200MM camera on a 40-cm telescope (JHT) at Bayfordbury. This demonstrates that, as the DR is higher on a 16-bit camera than on a 12-bit camera for a fixed exposure, multiple faint stars were detected with the 16-bit camera that were not visible with the 12-bit camera. On the other hand, long exposures cause oversaturation of the bright star, leading to non-linearity and charge spill-over into neighbouring pixels. This blooming effect negatively impacts the overall observation, making it impossible to measure reliable magnitudes and detect faint objects. These mutually exclusive scenarios can be compensated by the complex mode of observation, as displayed in the top-right panels of Figs 16 and 17.

The reconstructed images were then used for astrometric calibration to convert pixel coordinates to the world coordinate system using Astrometry.net[4] (Lang et al. 2010). Gaia DR2 Catalogue (Gaia Collaboration et al. 2018) was subsequently downloaded to identify detected stars in the field. We marked several faint stars of Gaia magnitude around 15 on both Figs 16 and 17, as they are clearly visible in our images.

Furthermore, we used 'DAOStarFinder' from the 'Photutils' Python library to detect stars in our images. 'DAOStarFinder' operates based on the 'DAOFIND' algorithm (Stetson 1987). We found that all relatively faint stars marked in Figs 16 and 17 could be detected with a detection threshold of $5\sigma$, where $\sigma$ is the standard deviation calculated using sigma-clipped statistics on the image data.

In addition to the relatively faint stars identified from the Gaia catalogue as mentioned above, we selected a few more stars to calculate photometric magnitudes from our complex readout mode observations. Following Stetson (1987), instrumental magnitudes for Vega and Polaris were first derived using aperture photometry using CircularAperture from Python's photutils package, based on observations in the standard readout mode. Since no typical photometric standard star field was observed, Vega was assumed to have a standard magnitude of 0.0 and Polaris a standard magnitude of 1.0 in our photometric system. These assumptions are reasonable, as Vega has I= 0.1 mag and J= −0.18 mag, while Polaris has I= 1.22 mag and J= 0.8 mag. These reference values were then used to determine the zero points for calculating the magnitudes of stars in the corresponding Vega and Polaris fields observed in the complex readout mode, as tabulated in Table 1. For our measurements, a circular aperture of 3 × FWHM was used, and the local background around each source was estimated using an annular aperture of a width of 1 × FWHM outside the star's aperture. The photometric errors for our measurements, as listed in Table 1, are $1\sigma$ background-only errors calculated using the 'calculate_magnitude_error' task in photutils.

We compared the magnitudes of our selected stars from various catalogues, including Gaia DR2, the Pan-STARRS1 Surveys (PS1-DR1; Chambers et al. 2016), and the 2MASS All-Sky Catalog of Point Sources (Cutri et al. 2003), as listed in Table 1. Cross-matching

[4] https://astrometry.net/

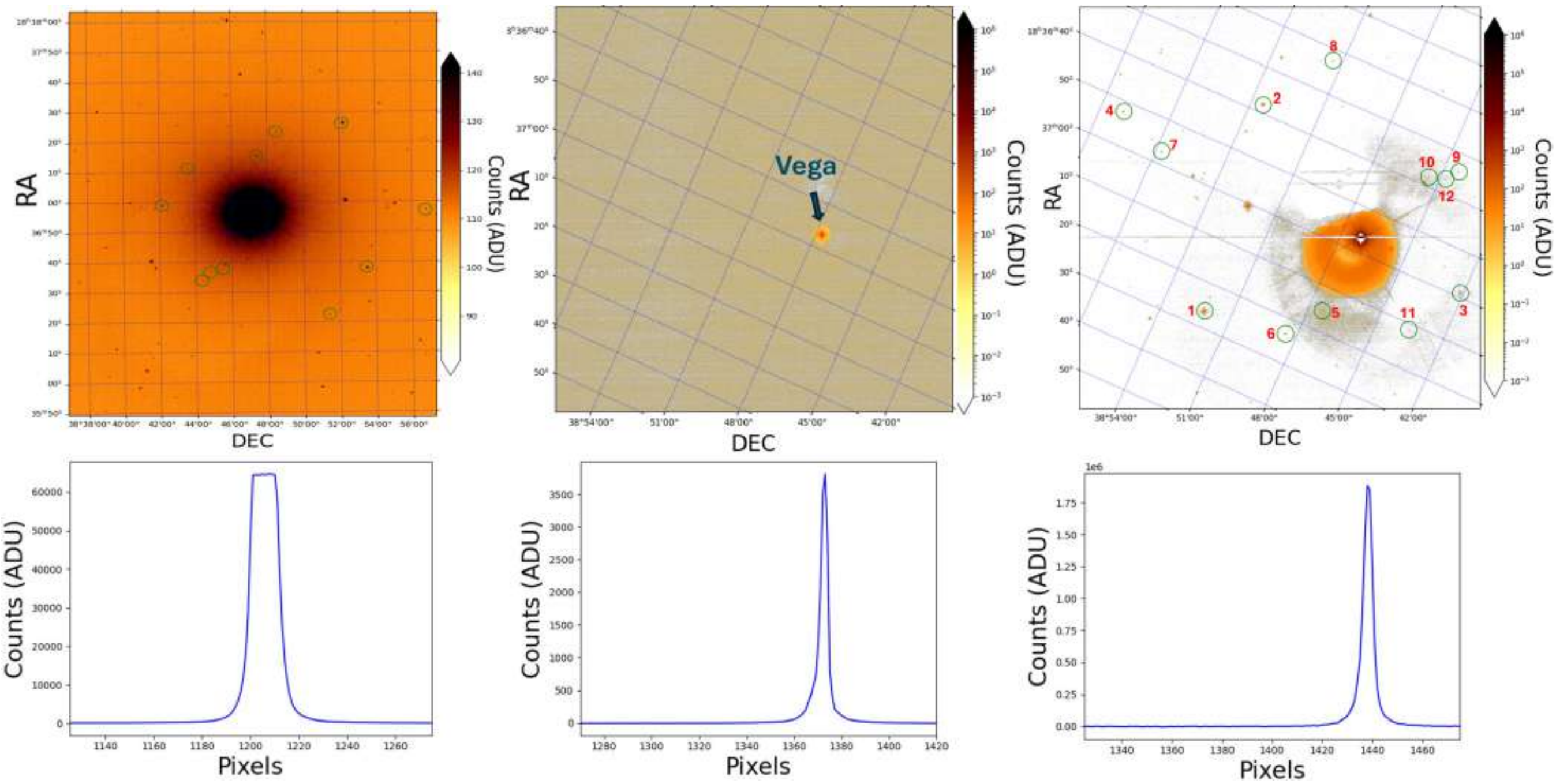

**Figure 16.** The top panels present a pictorial comparison of images obtained for Vega with a 16-bit camera (left), our standard read-out mode (middle), and the reconstructed image in the complex read-out mode (right), showing the differences in image contrast within a single frame. The bottom panels illustrate a cut along a row passing through the centroid of Vega in each respective image, demonstrating a 'flat-top' in the counts of an oversaturated image (left) and different unsaturated scenarios (middle and right). A few stars detected only in the complex mode are marked in green, and their magnitudes are provided in Table 1. These same stars are also marked in the left image, demonstrating that the 12-bit camera in complex mode is capable of detecting stars as faint as those detected by a 16-bit camera in standard mode, without saturating the bright object. The colour bar represents counts. The scale of the left-image ranges from m $-$ $10\sigma$ to m + $10\sigma$, where m and $\sigma$ are the sigma clipped median and standard deviation, respectively. The colour bars in the middle and right images are displayed on a logarithmic scale.

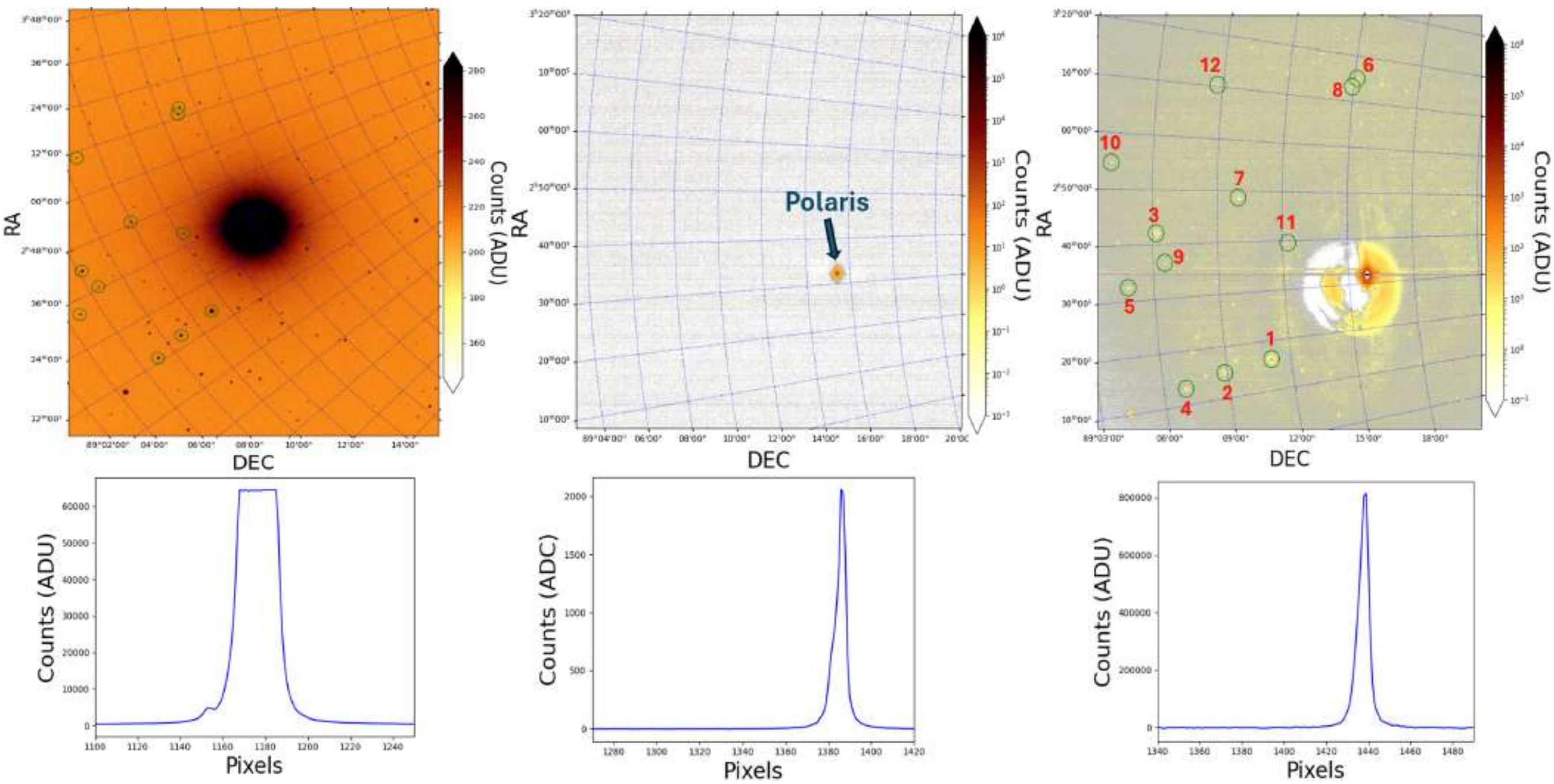

**Figure 17.** Similar to Fig. 16, but for Polaris.

**Table 1.** Details on identified stars.

| Star no | Gaia RA (degree) | Gaia DEC (degree) | G (mag) | BP − RP (mag) | $z_{SDSS}$ (mag) | $y'_{Baader}$ (mag) | $J_{2MASS}$ (mag) |
|---|---|---|---|---|---|---|---|
| Vega field: | | | | | | | |
| 1 | 279.359 | 38.869 | 9.95 ± 0.01 | 0.96 | 9.70 ± 999 | 9.85 ± 0.01 | 8.75 ± 0.03 |
| 2 | 279.159 | 38.891 | 10.57 ± 0.01 | 0.81 | 10.55 ± 0.05 | 10.78 ± 0.02 | 9.53 ± 0.02 |
| 3 | 279.246 | 38.702 | 11.98 ± 0.01 | 1.23 | 11.47 ± 999 | 11.90 ± 0.03 | 10.40 ± 0.02 |
| 4 | 279.218 | 38.983 | 12.11 ± 0.01 | 0.75 | 11.99 ± 999 | 13.05 ± 0.04 | 11.14 ± 0.02 |
| 5 | 279.314 | 38.790 | 12.60 ± 0.01 | 0.71 | 12.53 ± 0.02 | 12.77 ± 0.03 | 11.74 ± 999 |
| 6 | 279.348 | 38.808 | 12.87 ± 0.01 | 0.80 | 12.70 ± 0.01 | 13.71 ± 0.06 | 11.84 ± 0.02 |
| 7 | 279.238 | 38.946 | 13.02 ± 0.01 | 0.75 | 12.90 ± 0.01 | 14.08 ± 0.07 | 12.04 ± 0.02 |
| 8 | 279.094 | 38.857 | 13.17 ± 0.01 | 0.72 | 13.06 ± 0.01 | 14.30 ± 0.08 | 12.23 ± 0.02 |
| 9 | 279.142 | 38.739 | 13.63 ± 0.01 | 0.70 | 13.50 ± 0.01 | 14.68 ± 0.09 | 12.67 ± 0.02 |
| 10 | 279.158 | 38.758 | 14.77 ± 0.01 | 0.96 | 14.23 ± 0.01 | 15.51 ± 0.08 | 13.24 ± 0.02 |
| 11 | 279.297 | 38.726 | 15.02 ± 0.01 | 0.67 | 14.76 ± 0.01 | 15.81 ± 0.13 | 13.83 ± 0.04 |
| 12 | 279.153 | 38.746 | 15.11 ± 0.01 | 0.69 | 14.80 ± 0.01 | 15.99 ± 0.12 | 13.89 ± 0.02 |
| Polaris field: | | | | | | | |
| 1 | 34.083 | 89.185 | 10.61 ± 0.01 | 1.47 | 9.99 ± 999 | 9.85 ± 0.01 | 8.78 ± 0.03 |
| 2 | 33.733 | 89.148 | 12.07 ± 0.01 | 1.05 | 11.82 ± 999 | 11.80 ± 0.02 | 10.77 ± 0.02 |
| 3 | 40.523 | 89.103 | 12.22 ± 0.01 | 1.67 | 11.34 ± 999 | 11.31 ± 0.01 | 10.14 ± 0.02 |
| 4 | 33.271 | 89.117 | 12.38 ± 0.01 | 1.62 | 11.56 ± 999 | 11.81 ± 0.02 | 10.37 ± 0.02 |
| 5 | 38.122 | 89.080 | 13.03 ± 0.01 | 1.20 | 12.56 ± 0.01 | 12.75 ± 0.02 | 11.57 ± 0.02 |
| 6 | 48.787 | 89.253 | 13.47 ± 0.01 | 1.08 | 13.06 ± 0.01 | 13.14 ± 0.03 | 12.11 ± 0.02 |
| 7 | 42.187 | 89.166 | 13.53 ± 0.01 | 1.57 | 12.71 ± 0.01 | 12.75 ± 0.02 | 11.55 ± 0.02 |
| 8 | 48.308 | 89.249 | 13.60 ± 0.01 | 1.28 | 13.05 ± 0.01 | 13.07 ± 0.03 | 12.04 ± 0.02 |
| 9 | 39.154 | 89.109 | 14.80 ± 0.01 | 1.25 | 14.21 ± 0.01 | 14.81 ± 0.04 | 13.25 ± 0.03 |
| 10 | 43.684 | 89.068 | 14.85 ± 0.01 | 1.11 | 14.38 ± 0.01 | 14.65 ± 0.04 | 13.45 ± 0.03 |
| 11 | 39.887 | 89.204 | 15.25 ± 0.01 | 1.32 | 14.55 ± 0.02 | 14.55 ± 0.04 | 13.55 ± 0.03 |
| 12 | 47.589 | 89.146 | 15.14 ± 0.01 | 1.18 | 14.65 ± 0.01 | 14.74 ± 0.04 | 13.64 ± 0.03 |

Notes: In photometric error measurements, a value of 999 is mentioned to indicate missing, non-existent, or unreliable data.

and comparison between the Gaia DR2 and 2MASS catalogues were performed using SAOImageDS9. For PS1-DR1, the brightest stars were identified within a 1 arcminute radius of the given RA and DEC, having magnitudes between those of the G and J filters. If multiple stars were found with magnitudes between the G and J filters, the star closest to the given RA and DEC was selected for comparison. When comparing our measurements with PS1-DR1, we found good agreement for the Polaris field, with a mean difference of 0.14 mag. In contrast, a significant mean difference of 0.85 mag was observed for the Vega field. This discrepancy may be attributed to the high background caused by scattered light (see Section 5 for details) in case of Vega.

In addition, a row passing through the center of the brightest object was extracted for all three cases discussed above for Vega and Polaris. This comparison is illustrated in the bottom panel of Figs. 16 and 17. It can be seen that the left image was completely saturated for both Vega and Polaris. The middle one for Polaris was unsaturated, while Vega began to show saturation. The right image, which is read out in complex mode, is unsaturated for both Vega and Polaris. Therefore, both bright (near zero magnitude) and faint (approximately 15 magnitude) stars are detectable without saturation in a single exposure using our complex readout scheme, corresponding to an intensity ratio of about one million to one.

## 5 DISCUSSION

In this paper, we successfully demonstrate how the DR of a commercially available CIS120 image sensor can be enhanced. In addition, we provide a characterization of the sensor. As this is a development project, various experiments were conducted at different times, with different gain settings, and using different software versions, resulting in some lack of self-consistency. We discuss below the experimental trade-offs that we made in this study.

In laboratory experiments, we observed that for a given exposure time $T$ and equal illumination, the recorded counts in standard mode with a sensor size of 2K × 2K were about four times higher than in the complex mode, where 2 rows (to oversample the target PSFs, as mentioned earlier) are read out 4095 times. The determination of the actual exposure depends on various factors. For example, the read-out and reset times are a function of the choice, as well as the number of bright rows, and the total number of read-out. To avoid complexity and unwanted software issues while observing during evolving sky conditions, the exposure times for on-sky experiments were not set below 20 ms in both standard and complex read-out modes.

Observations of 'Vega', in complex mode, as the recorded counts were approximately four times lower than in standard mode, the exposure time was set four times greater than in the standard mode to maximize the dynamic range of the sensor. However, observations of 'Polaris', we used the same exposure time for both standard and complex modes. We did not increase the exposure time by a factor of four in complex mode because this would have significantly increased observation time, and we did not want to push the tracking capability of the auto-guider. As a result, for 'Polaris', the DR of the sensor was not maximized.

To enhance the DR, we aimed to maximize the number of times the bright rows were readout, while minimizing the number of 'bright' rows for multiple readouts. We also needed to ensure that the centroid of the image was captured, and, given the relatively large plate scale, atmospheric turbulence led us to compromise at 10 pixels. This is demonstrated in Fig. 15. The count level in each row also fluctuated with each readout. As a result, when we stacked all individual readouts of the bright rows, summed them, and selected the brightest row for extraction, the selected row may not have contained the maximum counts in each individual read-out. This is apparent in our results (see Figs 16 and 17), where the limitations discussed above can be clearly observed.

Moreover, the optical setup used for our laboratory experiments was not optimized for DR. In particular, we did not investigate the optimal exposure needed to achieve maximum counts from the bright row without saturation and minimum counts from the faint row just above the sensor's detectable threshold. Furthermore, the laboratory housing and cooling system, constructed from stainless steel, together with camera window and the uncoated glass window of the chip, resulted in high internal reflection. As a result, the experiment had to be conducted carefully to observe the first and third transmitted images, as mentioned earlier. Nevertheless, the removal of scattered light was not perfect, as the blocking of the third beam to obtain the scatter frame was not optimized. Thus, the faint object and its background included additional counts due to scattered light.

In addition, we observed the same scattering effect in the telescope housing system during on-sky experiments, likely originating from the diverging beam reflected off the uncoated glass window on the chip, the camera window and the filter used. This effect was more severe in observations of Vega, where the DR was maximized as described above. Due to this effect, ghost images was formed around the bright object and a large, disk-like feature was observed surrounding it. Typical sky subtraction using the five-point dithering method for generating the near-infrared sky during data analysis was not sufficient to eliminate this artifact. As a result, we were unable to detect any faint objects close to the bright object. We believe that this effect can be mitigated by using a windowless chip with an improved housing system. A physical mask placed over the bright object could be another solution.

## 6 SUMMARY

In this work, we investigated the performance of a commercial CMOS image sensor, the Te2v CIS120. Our primary motivation was to enhance the dynamic range of CIS120 while providing a comprehensive characterization of the sensor. We developed two housing systems for the sensor: one for laboratory experiments and another for on-sky observations. In the laboratory setup, we utilized a two-stage thermoelectric cooler with a liquid chiller and a vacuum chamber. For the telescope housing system, we employed a three-stage thermoelectric cooler with a liquid chiller and circulated dry nitrogen to prevent condensation. We first measured the bias level, read noise, dark current, linearity, and system gain using the laboratory housing. At a sensor temperature of $-10°$C and low-gain setting, the mean bias and mean read noise were found to be 2652 $e^-$ and 10.2 $e^-$, respectively. The mean dark current in high-gain setting reaches a value of 0.004 $e^-$/s at $-50°$C. At the same temperature, the measured system gain was 1.77 $e^-$/ADU.

To enhance dynamic range, we implemented row-readout sequencing, in which the sensor's rows are divided into two groups: one group is read out multiple times, while the other is read out only once to form the final image. We demonstrated our dynamical enhancement scheme in both laboratory and on-sky experiments. In the laboratory, we achieved a dynamic range enhancement of 63 dB for a 12-bit sensor. For on-sky experiments, we observed Vega and Polaris fields, which contain both bright stars (with magnitudes around 0) and faint stars. Our scheme enabled us to detect stars as faint as 15th magnitude with a $5\sigma$ detection threshold without saturating the bright stars near zero magnitude, corresponding to a dynamic range of approximately one million to one for a 12-bit astronomical camera. It is possible to further enhance the on-sky dynamic range by using less bright readout rows and with a more careful choice of exposure times and experimental setup. In principle, it is also possible to modify the detector readout to enable an increased number of row reads.

## DATA AVAILABILITY

The data underlying this article will be shared upon reasonable request to the corresponding author.

## ACKNOWLEDGMENTS

The authors are thankful to the reviewers for their critical and valuable comments on the original version of the manuscript, which helped us to improve the paper. This research was made possible thanks to UKSA grant ETP2-037. Teledyne e2v generously provided the CIS120 used for this experiment. We are indebted to Dave Colebook and Karen Holland at XCAM for their great patience in dealing with the firmware modifications. We are grateful to Eugen Guritanu for design and printing of the camera housing, to Michael Wilkins for machine shop fabrication, and Andrew Carruthers at Finetech for the fabrication. Observations at Bayfordbury Observatory were possible thanks to Sam Rolfe for their support with experimental setup. Laboratory support was provided by Krupa Kuruvila. This research also benefit from high performance computing infrastructure ST/R000905/1 provided by the STFC to the University of Hertfordshire.

## CONFLICTS OF INTEREST

Authors declare no conflict of interest.